%% file: main.tex
\documentclass[letterpaper,twocolumn,10pt]{article}
\usepackage{usenix-2020-09}

\usepackage{tikz}
\usepackage{amsmath}

\usepackage{filecontents}

\usepackage{comment}
\usepackage{balance}
\usepackage{subcaption}
\usepackage{enumitem}
\usepackage{xurl}  

\usepackage{booktabs}

\usepackage{tabularx}
\usepackage{graphicx}
\usepackage{multirow}
\usepackage{array}
\usepackage{adjustbox}
\usepackage{lscape} 
\usepackage{rotating}
\usepackage{siunitx}

\usepackage{adjustbox}

\usepackage{ltablex}
\usepackage{caption}
\usepackage{multicol}
\usepackage{footnote}
\usepackage{colortbl}
\usepackage{longtable}
\usepackage{float}
\usepackage{textcomp}       
\usepackage{amssymb} 
\DeclareUnicodeCharacter{202F}{\,}     
\DeclareUnicodeCharacter{03B5}{$\varepsilon$} 
\DeclareUnicodeCharacter{2264}{$\leq$}
\DeclareUnicodeCharacter{2265}{$\geq$} 
\DeclareUnicodeCharacter{2113}{$\ell$} 
\DeclareUnicodeCharacter{221E}{$\infty$} 
\DeclareUnicodeCharacter{0394}{$\Delta$} 
\DeclareUnicodeCharacter{2248}{\ensuremath{\approx}}

\definecolor{headercolor}{RGB}{230,230,230}
\definecolor{layercolor}{RGB}{245,245,245}
\newcolumntype{L}[1]{>{\raggedright\arraybackslash}p{#1}}
\newcolumntype{C}[1]{>{\centering\arraybackslash}p{#1}}
\newcolumntype{R}[1]{>{\raggedleft\arraybackslash}p{#1}}
\newcommand{\CRR}{\operatorname{CRR}}
\newcommand{\CCI}{\operatorname{CCI}}
\newcommand{\clip}[1]{\operatorname{clip}\!\left(#1\right)}

\begin{document}
\date{}

\title{SoK: Cybersecurity Assessment of Humanoid Ecosystem}

\author{
{\rm Priyanka Prakash Surve, Asaf Shabtai, Yuval Elovici}\\
Ben-Gurion University of the Negev
}

\maketitle

\begin{abstract}
Humanoids are progressing toward practical deployment across healthcare, industrial, defense, and service sectors. 
While typically considered cyber-physical systems (CPSs), their dependence on traditional networked software stacks (e.g., Linux operating systems), robot operating system (ROS) middleware, and over-the-air update channels, creates a distinct security profile exposing them to vulnerabilities that conventional CPS models do not fully capture.     
Prior studies have mainly examined specific threats, such as LiDAR spoofing or adversarial machine learning (AML). 
This narrow focus overlooks how an attack targeting one component can cascade harm throughout the robot's interconnected systems.
We address this gap through a systematization of knowledge (SoK) that takes a comprehensive approach, consolidating fragmented research from robotics, CPS, and network security domains. 
We introduce a seven-layer security model for humanoid robots, organizing 39 known attacks and 35 defenses across the humanoid ecosystem-from hardware to human-robot interaction. 
Building on this security model, we develop a quantitative 39×35 attack–defense matrix with risk-weighted scoring, validated through Monte Carlo analysis. 
We demonstrate our method by evaluating three real-world robots: Pepper, G1 EDU, and Digit.
The scoring analysis revealed varying security maturity levels ranging from 39.9\% to 79.5\% across the platforms.
This work introduces a structured, evidence-based assessment method that enables systematic security evaluation, supports cross-platform benchmarking, and guides prioritization of security investments in humanoid robotics.
\end{abstract}

\input{introduction}

\input{related_work}

\input{architecture}
\input{attack_surface_threat_landscape}
\input{gap_analysis}

\input{conclusion}
\clearpage
\appendix

\bibliographystyle{plain}

\bibliography{references}

\input{appendix}

\end{document}

%% file: introduction.tex
\section{Introduction}
\label{sec:introduction}

\textbf{Humanoids.} Humanoids are being deployed in sectors such as healthcare, industrial automation, defense, and service sectors by replicating human motion, cognition, and interaction~\cite{sutikno2024future,malik2023intelligent,anon2025humanoid}. 
Powered by AI, these robots are designed for real-time, high-stakes tasks such as surgical assistance, elder care, and public interaction. They also enable close collaboration in dynamic human environments~\cite{zhang2024perceptive}. 
However, the very capabilities that make humanoids so promising also expose them to new cybersecurity risks. Before these systems are widely deployed, it is critical to establish a rigorous method for assessing their security.
To address this challenge, we present a comprehensive systematization of knowledge (SoK) that introduces a seven-layer security architecture model specifically designed for humanoid robots. 
With the help of this model, we organize 39 documented attacks and 35 defense mechanisms across the humanoid technology stack, from hardware components to human-robot interaction layers. 
The purpose of this security model is to enable systematic vulnerability assessment, support cross-platform security benchmarking, and guide evidence-based security investment decisions for humanoid deployments.

\noindent\textbf{A unique and layered threat landscape.}
The security challenges faced by humanoids are fundamentally different from those of classical CPSs or traditional IT systems. Humanoids are perimeter-less assets: they work outside secured facilities and can be approached directly by humans in hospitals, factories, and public venues. This reality elevates the importance of Physical and Social interaction and reshapes threat priorities beyond traditional IT/CPS assumptions.
Humanoids must navigate dynamic, unpredictable settings with constantly shifting variables like human movement and environmental obstacles~\cite{verma2025systematic}.
Their integration into human-centered environments introduces a unique threat landscape, where cyber attacks on humanoid systems can jeopardize user safety, compromise privacy, and disrupt operational reliability~\cite{kinzler2019humanoid,yousef2024darwinop2}. 
These security breaches may lead to direct and immediate physical consequences.

This threat landscape is exacerbated by the deeply interconnected nature of humanoid architecture, where multi-layered subsystems create cross-layer dependencies and an expansive attack surface~\cite{Portugal2017,Nazari2023,Munir2023,Yaacoub2021,Elikhchi2023}. 
Unlike traditional CPSs, humanoids integrate numerous attack-prone subsystems: AI accelerators, sensor arrays, middleware, and decision-making algorithms-each with distinct vulnerability profiles that can cascade failures across the entire system.
At the \emph{hardware level}, humanoids face firmware tampering and sensor spoofing, with their AI accelerators (e.g., Jetson Orin, Neural Processing Units (NPUs)) being potential targets for exploits that compromise system integrity~\cite{Staffa2018}. 
In the \emph{decision-making area}, their reliance on deep learning for navigation and decision-making makes them vulnerable to adversarial and data poisoning attacks, which can trigger unpredictable or dangerous behaviors~\cite{Mazzeo2020,Patel2022}. 
Finally, their \emph{communication and middleware infrastructure}, often based on ROS 2, provides openings for man-in-the-middle attacks, unauthorized control hijacking, and real-time data manipulation~\cite{Deng2022,Jeong2017}. Mitigating these varied, layered threats requires a comprehensive and specialized security approach.
    
\noindent\textbf{Vulnerabilities in real-world humanoid platforms.}
While no major, publicly documented cyberattacks have targeted humanoids to date, the vulnerabilities described above are not merely theoretical. 
These vulnerabilities are embedded in the designs of humanoids, both in development and already deployed; hence proactive security assessment is required~\cite{Elikhchi2023,Priyadarshini2017,RodriguezLera2017,Botta2023}.

For instance, \emph{Tesla's Optimus}, designed for industrial automation, leverages AI-based motion planning. 
This design choice exposes it to adversarial AI manipulation, where carefully crafted malicious inputs could compromise control, disrupt manufacturing processes, or cause unsafe physical actions~\cite{Botta2023}. 
Evidence from AML research suggests potential susceptibility of Optimus’s AI-driven control to adversarial perturbations~\cite{Cao2023,pan2022}.

\emph{SoftBank's Pepper}, a widely deployed social humanoid, relies on networked connectivity and AI-powered facial recognition. 
This creates a significant privacy risk. 
Network vulnerabilities can be exploited for unauthorized data exfiltration and surveillance, while biased or manipulated AI could lead to unnatural or inappropriate social interactions~\cite{Mercuri2022}. 
This underscores the need for robust encryption and access control to protect the sensitive user data these platforms handle~\cite{RodriguezLera2017}.

Similarly, \emph{Unitree Robotics'} humanoid models utilize ROS-based middleware in industrial and tactical applications. 
This theoretically exposes them to a well-known class of exploits, which includes command injection, adversarial perception, and man-in-the-middle attacks that are common in insecure ROS deployments~\cite{Elikhchi2023,Botta2023}. 
Securing these middleware protocols remains a critical and ongoing research priority~\cite{RodriguezLera2017}.

The examples highlight the fact that modern humanoid architectures are susceptible to the security vulnerabilities discussed earlier.
As these robots become increasingly autonomous and integrated into our daily lives, a systematic approach for identifying and mitigating these risks is essential for ensuring humanoids' safe and trustworthy operation.

%% file: related_work.tex
\section{Related Work and Research Gaps}
\label{sec:related_work}
Humanoid security research spans robotics, AML, CPSs, and human-computer interaction domains. 
However, existing research remains fragmented across these disciplines, lacking unified architectural models and systematic analyses of cross-layer vulnerability propagation. 
Most prior work has focused on isolated threats without considering how attacks targeting one subsystem can cascade throughout the interconnected humanoid architecture.
Our SoK addresses this fragmentation by conducting systematic literature analysis to answer three key research questions: (1) What are the primary security threats facing humanoid robots and how do they relate to its system architecture? (2) What gaps exist in current attack-defense coverage? (3) Can architectural patterns from the literature help in systematic security assessment approaches for humanoid platforms?
Our analysis synthesizes 89 high-confidence studies to reveal consistent architectural patterns that inform the development of a structured security model specifically designed for humanoid systems.

\noindent\textbf{Research methodology and paper selection. }
We conducted a PRISMA-compliant systematic review targeting humanoid security research across four major academic databases (IEEE Xplore, ACM Digital Library, SpringerLink, and arXiv) using the search keywords: ("humanoid robot" OR "social robot" OR "service robot") AND ("security" OR "cybersecurity" OR "adversarial" OR "attack"), spanning publications from 2015-2025.
Our inclusion criteria required: (1) explicitly address security with concrete threat analyses, (2) embodied robotics relevance beyond theoretical security models, and (3) peer-reviewed publications or high-quality preprints. 
Studies were excluded if they focused solely on theoretical security concepts without application to robotics, or if they pertained to industrial automation lacking humanoid characteristics. 
The initial search yielded 530 papers. 
After title/abstract screening, 150 papers underwent full-text analysis, resulting in 89 high-confidence studies that form our research corpus.
Through systematic content analysis, we identified 39 distinct attack vectors and 35 defense mechanisms, which clustered around seven functional architectural layers. 
This corpus shows rapid growth, with 68\% published after 2020, indicating rapid field emergence and the need for systematic organization of this growing knowledge base.

\noindent \textbf{Existing Security Taxonomies and Their Limitations. }
Prior research has approached robotics security through various taxonomic frameworks, each providing valuable but incomplete perspectives on the threat landscape.
Traditional IT security models, while comprehensive within their domains, prove inadequate for embodied autonomous systems. 
For example, established vulnerability surveys identify 47 distinct vulnerability classes across hardware, communication, and software domains, but these classifications fail to capture the unique characteristics of embodied AI systems where physical actions can result from compromised digital processes~\cite{Yaacoub2021}. 

The coupling of cyber and physical domains in humanoids creates attack vectors that traditional IT security models do not address.
Cyber-physical systems (CPS) security frameworks provide better alignment with robotics contexts but remain focused on industrial control systems rather than autonomous humanoids~\cite{Wang2021}. 
These models excel at analyzing control loop vulnerabilities but lack the architectural depth needed to address AI-driven decision-making, dynamic human interaction, and autonomous mobility that characterize modern humanoids.
Robot-specific security analyses have made significant empirical contributions, including comprehensive assessments of ROS-based systems that identified 114 security issues across 15 robotic platforms through red-team approaches~\cite{Mayoral2020}. 

While demonstrating practical exploitability, these platform-specific studies lack the architectural abstraction necessary for systematic cross-platform analysis and fail to model how vulnerabilities propagate across the layered subsystems that comprise integrated humanoid architectures.
Our review confirms that most studies examine isolated threats (sensor spoofing, adversarial ML) rather than modeling how they cascade across architectures. 
This gap motivates our development of a layered architectural model that organizes security analysis around the functional boundaries inherent to embodied autonomous systems.

\subsection{Current State: Siloed Security Approaches}
Our systematic analysis of 89 high-confidence papers reveals a research landscape characterized by domain-specific depth but limited cross-domain integration. 
While individual security domains have developed sophisticated attack-defense frameworks; they remain bounded by traditional domains, limiting their ability to capture cross-layer propagation in understanding system-wide vulnerabilities.

\noindent \textbf{AML in Robotics. }
The intersection of AML and robotics has emerged as an active research area, though coverage remains narrow. 
Recent work demonstrates sophisticated attacks against core AI components: visual simultaneous localization and mapping (SLAM) systems have been compromised through carefully crafted perturbations that cause localization failures in navigation tasks, while reinforcement learning policies have been manipulated to produce suboptimal or dangerous actions during critical decision-making phases~\cite{Chen2024AoR,Buddareddygari2022RLSign}. 
Sequential perturbation attacks against quadrupedal locomotion systems have shown how adversaries can induce gait instabilities that compromise mobility, while supply chain attacks have been demonstrated against vision-language models used for robotic manipulation tasks~\cite{Shi2024ANYmal,wang2024trojanrobot}. 
These studies indicated that while attack methodologies are well-documented across the surveyed papers, corresponding defense mechanisms appear in a few, creating a significant vulnerability gap. 
Moreover, these studies focus exclusively on isolated AI components without considering how AML attacks might propagate across integrated robotic architectures.

\noindent \textbf{Middleware and Communication Security. }
Within the reviewed papers, ROS and data distribution service (DDS) security were among the most studied topics, with approximately one-third of the surveyed publications addressing both attack vectors and corresponding defense mechanisms~\cite{Deng2022,Rivera2019,Antunes2022}. 
Researchers have found that ROS and DDS are vulnerable to attacks like unauthorized command injection~\cite{Sugawara2020LightCommands,Zhang2017DolphinAttack}, and some defenses such as intrusion and anomaly detection~\cite{SorianoSalvador2024} have been proposed. However, the current research treats middleware as an isolated layer, failing to model how communication vulnerabilities interact with sensing, planning, and actuation subsystems in real-world deployments.

\noindent \textbf{Physical and Hardware Security. }
Hardware security research has systematically explored attack vectors and corresponding defenses, with strong emphasis on experimental validation~\cite{Quarta17}. 
Seminal work on industrial robot controllers demonstrated how firmware tampering and parameter manipulation can compromise fundamental safety properties, while cyber-physical attacks through additive manufacturing have shown how physical modifications can create persistent backdoors~\cite{Belikovetsky17,VacuumACPoster,Shah22}. 
Acoustic and electromagnetic side-channel attacks have been systematically explored, revealing how sensitive operational data can be extracted through passive monitoring. 
Defensive innovations include tamper-evident hardware designs and secure boot mechanisms that provide corresponding countermeasures~\cite{Vidakovic23}. 
However, hardware security research typically focused on individual device vulnerabilities without considering how hardware compromises might facilitate attacks across integrated system architectures.

\noindent \textbf{Sensor and Perception Security. }
Research on sensor security reflects a balanced development of attack and defense strategies. 
LiDAR spoofing attacks have evolved from simple laser interference to sophisticated geometric manipulation that can induce arbitrary depth perceptions~\cite{Cao2023,Shin17}, while adversarial patch attacks against object detection have progressed from laboratory demonstrations to real-world feasibility studies~\cite{Thys19,Eykholt18}. 
Camera-based attacks now include both digital perturbations and physical modifications that remain effective across varying environmental conditions~\cite{Petit2015}. 
Defensive research has responded with sensor fusion approaches, coded illumination systems, and temporal consistency checks that provide robust countermeasures~\cite{wang2015,You2021}. 
However, defenses are typically evaluated against individual attack types rather than coordinated multi-sensor attacks that could overwhelm detection mechanisms.

\noindent \textbf{Human-Robot Interface Security. }
Human-robot interaction security represents the most underdeveloped domain, appearing in a few surveyed papers with limited systematic coverage. Acoustic attacks have demonstrated how inaudible command injection can bypass voice authentication systems, while adversarial audio embedding techniques can hide malicious commands in seemingly benign communications~\cite{Zhang2017DolphinAttack,Yuan2018CommanderSong}. Social engineering approaches exploit trust relationships between humans and robots, manipulating user behavior through deceptive interactions~\cite{Aroyo2018TrustSE,Belpaeme2019RobotPersuasion}. Recent work has examined multimodal attacks against depth-based authentication systems, revealing vulnerabilities in biometric security mechanisms~\cite{Wu2023DepthFake}. This domain lacks comprehensive defensive frameworks, with most proposed countermeasures addressing individual attack vectors rather than systematic interface security.

\subsection{Recurring Architectural Themes}
\label{subsec:pattern_recognition}
Our systematic analysis uncovered a distinct pattern: despite the apparent diversity of humanoid platforms and attack methodologies examined across papers, security concerns consistently clustered around seven distinct functional boundaries. 
During our content analysis, threats that initially appeared unrelated; consistently aligned with the same architectural layers.
For example, sensor-spoofing attacks, whether targeting LiDAR systems~\cite{Shin17,Cao2023} or cameras~\cite{Thys19,Eykholt18}, invariably exploited vulnerabilities at the sensory interface between physical hardware and digital processing. 
These attacks succeeded by manipulating the boundary where analog sensor data transforms into digital representations used by higher-level algorithms. Similarly, AML attacks consistently targeted high-level reasoning components, exploiting the interface between data processing and decision-making logic~\cite{Chen2024AoR,Buddareddygari2022RLSign,Shi2024ANYmal}. Physical components, upon which all other system functions depend, can be compromised by hardware attacks.
Most significantly, the observed patterns reflect real architectural boundaries in humanoid robots-not merely the areas researchers emphasized.
Unlike traditional IT systems with arbitrary software boundaries, humanoids have natural layers that reflect their physical and digital components. These layers create consistent attack points where different subsystems connect, making them common targets across different platforms.

\subsection{Critical Gaps}

Our analysis shows that robotics security studies intensely focus on specific attack vectors or individual subsystems; only a few of the surveyed papers examined attacks spanning more than one subsystem or architectural layer. 
As a result, prior work often fails to capture realistic adversarial dynamics, where a local fault can propagate across the humanoid layers.
We observe an asymmetric defense landscape. Human–robot interface (HRI) vulnerabilities are well documented~\cite{Zhang2017DolphinAttack,Yuan2018CommanderSong}, but practical, validated mitigations are scarce. By contrast, several sensor-level attacks (e.g., LiDAR spoofing, IMU drift)~\cite{Shin17,Cao2023} already have concrete countermeasures (filtering, fusion checks, shielding, freshness tests)~\cite{wang2015,You2021}. The result is an imbalanced posture: upper layers that shape user intent are weakly protected, while lower layers have spot-fixes.
Modern robotic systems need more measures like integrated defenses that work together across all system layers. 
This means binding human-robot interaction policies to authenticated, session-verified middleware; requiring sensor validation before any planning occurs; and implementing hard limits at the control level with automated recovery capabilities.
The proposed model makes this integration measurable by identifying the specific layers and interfaces that must connect, while RISK-MAP quantifies whether a platform actually implements these connections and calculates the residual risk when any link fails.

\subsection{Contributions of This Paper}

This work addresses the identified gaps through the following novel contributions:

\begin{enumerate}[left=0pt, nosep, leftmargin=*]
    \item \textbf{Comprehensive literature synthesis:} Systematization of the humanoid literature, integrating diversified cybersecurity findings across hardware, middleware/communications, machine learning, and human–robot interaction revealing critical coverage gaps and methodological limitations.    
    \item \textbf{Novel seven layered model:} Introduction of seven-layer security model tailored to humanoids that organizes 39 attack vectors and 35 defenses across natural system boundaries identified through our literature analysis.
    \item \textbf{Cross-layer vulnerability analysis:} Identification of cascading vulnerabilities and dependencies across multiple system layers, emphasizing integrated defense strategies rather than isolated countermeasures.
    \item \textbf{Quantitative assessment method:} We distill from the literature, a risk-weighted scoring system with statistical validation that enables systematic security evaluation and cross-platform benchmarking.
    \item \textbf{Empirical validation: }Demonstration of our method through evaluation of three real-world humanoid platforms, revealing specific vulnerability patterns and defense imbalances that inform practical security investment decisions.
\end{enumerate}

%% file: architecture.tex
\section{A Layered Security Architecture} 
\label{sec:architecture}

\subsection{Architectural Analysis}
\begin{figure}[h]
    \centering
    \includegraphics[width=\linewidth]{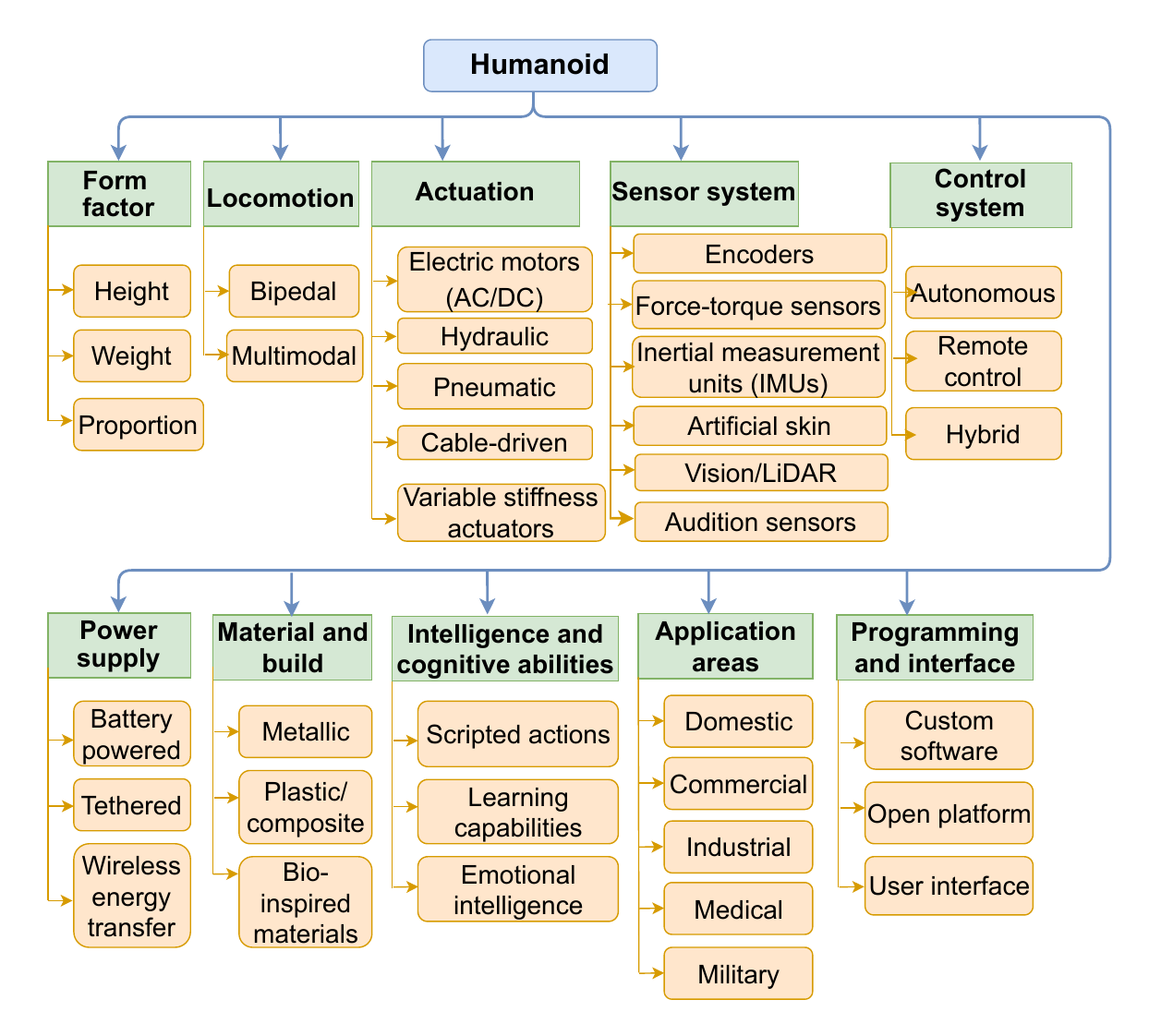}
    \caption{Taxonomy of humanoid designs.}
    \label{fig:humanoid-taxonomy}
\end{figure}

The recurring attack patterns identified in Section~\ref{sec:related_work} call for deeper architectural investigation. 
To understand why security vulnerabilities consistently cluster around specific functional boundaries, we examined the fundamental structure of humanoids across diverse platforms. Figure~\ref{fig:humanoid-taxonomy} organizes humanoid designs along ten design axes (morphology/actuation, power, mobility, sensing mix, compute, middleware, autonomy, interface modality, and deployment domain) which we use in our analysis. 

\noindent \textbf{The network-CPS duality. }
Our analysis uncovered that humanoids exist at the intersection of two established computational paradigms.
Like networked systems, they use layered abstractions with defined interfaces, protocols, and modular data flows. Like CPSs, they couple computation to physical processes under real-time and safety constraints. This duality implies that an effective security architecture must treat interfaces as enforcement points for identity, authorization, integrity, confidentiality, and auditability and practically, use each interface to enforce who can communicate, what can be sent, whether it is within safe bounds, and recorded for audit.

\noindent \textbf{Architectural convergence despite design diversity. }
From our review, we observed that across the surveyed platforms, security issues repeatedly surface at interface boundaries. 
This interface-centric clustering indicates an underlying architectural organization and motivates the seven-layer model we introduce.
These layers are: physical (responsible for actuation and mechanical structure), sensing and perception (handling physical-to-digital conversion and environmental awareness), data processing (managing real-time control loops and computational tasks), middleware (coordinating communication and data integration), decision-making (executing planning and reasoning), application (implementing task-specific behavior), social-interface (managing human-robot interaction).
This convergence explains why the attack patterns observed in Section~\ref{sec:related_work} concentrate at interfaces: the clustering reflects architectural boundaries in humanoid robots rather than researcher focus. 
This finding provides a strong basis for a generalized security model.

The identification of functional layers prompted us to examine established security models. While alternatives such as the TCP/IP stack and the Purdue Enterprise Reference Architecture (PERA) were considered, these frameworks focus primarily on communication protocols (TCP/IP) or industrial processes (PERA). In contrast, the Open Systems Interconnection (OSI) model provides a more general abstraction of layered dependencies, offering the closest conceptual alignment with the hierarchical structures observed in humanoid systems. OSI’s proven effectiveness in structuring complex security analyses and in capturing cross-layer dependencies made it a natural foundation for our seven-layer security model.

Therefore, we developed a seven-layer security architecture tailored to humanoids by refining the OSI model, in an iterative process. This process involved the following steps:

\begin{enumerate}[left=0pt, nosep, leftmargin=*]
    \item \textbf{Layer Identification:} Mapping the seven functional boundaries observed in our literature review (see Section~\ref{subsec:pattern_recognition}) to concrete subsystems across diverse humanoid platforms.
    \item \textbf{Threat Validation:} Verifying that the 39 distinct attack vectors identified in our analysis align with the identified functional boundaries.
    \item \textbf{Dependency Analysis:} Tracing how vulnerabilities can cascade across layers, reflecting the integrated nature of robotic systems.
    \item \textbf{Defense Integration:} Examining how the 35 defense mechanisms in the literature review can be systematically organized in the layered model.
\end{enumerate}
This iterative process resulted in the seven-layer model, which we formally define in the next subsection.

\subsection{Seven-Layer Humanoid Security Model}
\begin{figure}[h]
    \centering
\includegraphics[width=0.96\linewidth]{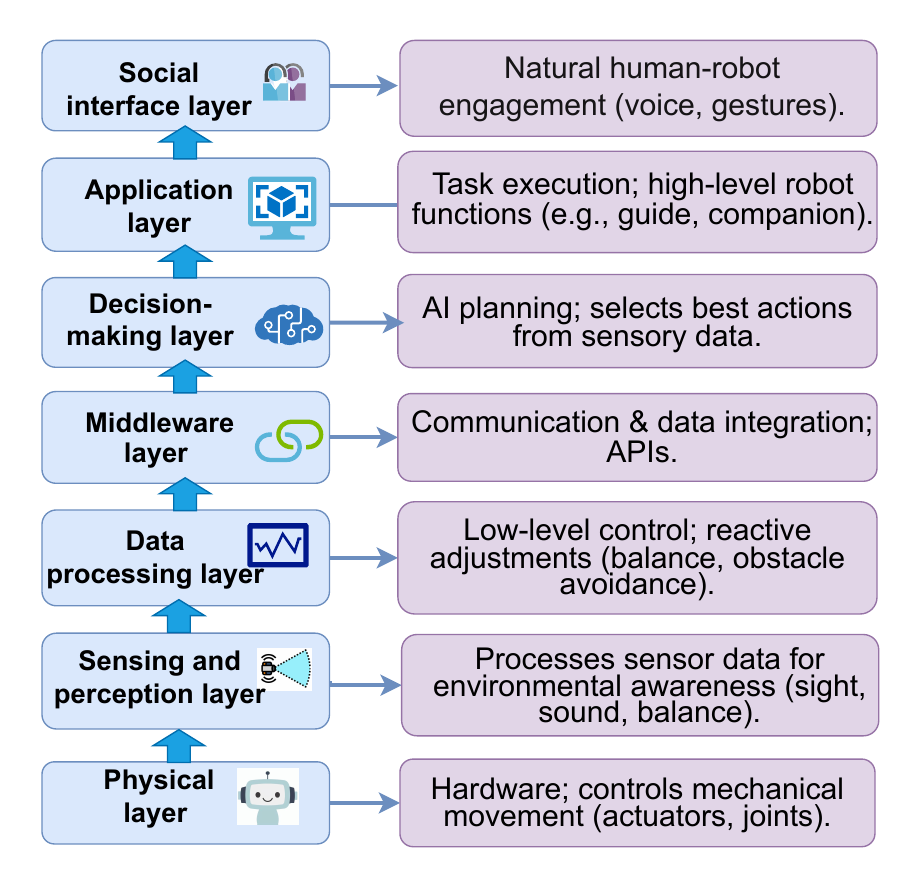}
    \caption{The seven-layer model for humanoids.}
    \label{fig:humanoid-layeredModel}
\end{figure}

\begin{figure}[h]
    \centering
    \includegraphics[width=\linewidth]{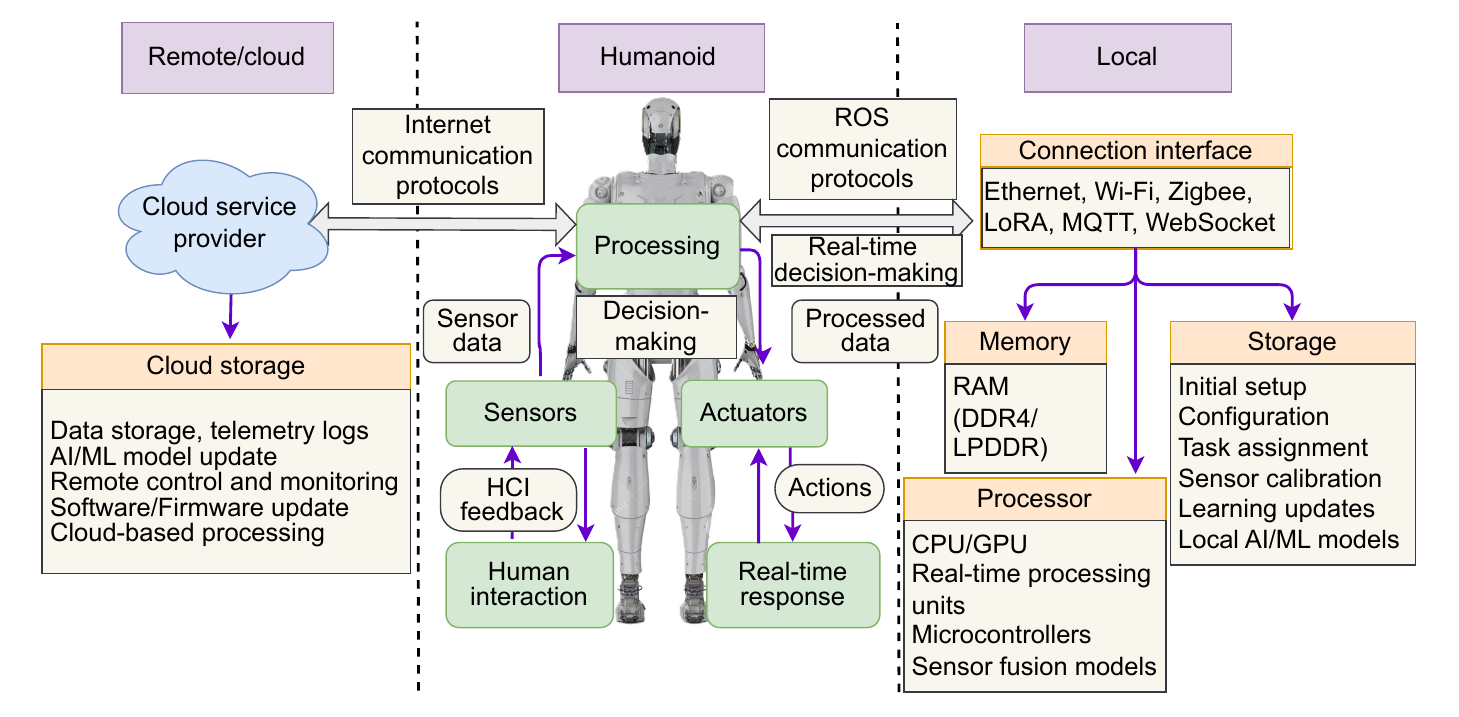}
    \caption{The humanoid in its cyber-physical ecosystem. 
    On-board sensing, processing, and actuation form a local feedback loop for real-time control. Cloud-based components support asynchronous tasks such as learning, telemetry, and Over-The-Air updates. 
    This view highlights that humanoids are nodes in broader cyber-physical networks.}
    \label{fig:humanoid-architecture}
\end{figure}

Our model structures humanoid architecture into seven distinct layers based on the functional boundaries identified in our analysis. Figure~\ref{fig:humanoid-layeredModel} defines these layers, which appear consistently across platforms and provide structure for systematic security analysis. 
This model directly reflects the operational reality of humanoids, as demonstrated in the cyber-physical ecosystem view in Figure~\ref{fig:humanoid-architecture}. 
The local sense-plan-act loop captures the real-time, embodied nature of these systems, while cloud integration reflects their dependence on distributed computational resources.


This progression from identifying literature fragmentation to recognizing architectural patterns and developing our layered model establishes the foundation for comprehensive threat analysis. In Section~\ref{sec:attack_surface}, we demonstrate how this model enables systematic analysis of both individual threats and their cascading effects across the humanoid ecosystem.

%% file: attack_surface_threat_landscape.tex
\section{From Ecosystem to Layer‑Wise Threat Landscape}
\label{sec:attack_surface}
\subsection{Holistic Humanoid Threat Landscape}

Humanoids' unique threat landscape stems from their dual nature: they inherit vulnerabilities from both conventional IT components  and their tightly integrated CPS elements. Our seven-layer model (see Section~\ref{sec:architecture}) enables structured security analysis by providing a systematic model for understanding risk distribution across system components. Using this model, we mapped  39 attacks and 35 defenses to specific architectural layers, as we have done in Tables~\ref{tab:attack_surface} and~\ref{tab:defences}.
This mapping demonstrates that threat patterns consistently align with architectural function, confirming the model's practical utility for identifying vulnerable components and understanding how attacks propagate across system boundaries.

Unlike conventional systems with loosely connected components, humanoids tightly integrate sensing, decision-making, and actuation in real-time control loops. This close coupling creates cascade effects, where a local compromise, such as a JTAG overwrite (P-A1), can bias state estimation (DP-A4) and manipulate high-level policies (DM-A5) without any additional network breach. Such vertical coupling expands the attack surface, allowing low-level faults to escalate into full-system compromise.

Real-time constraints make latency itself an attack vector. In enterprise IT, a few seconds of delay in anomaly detection may be tolerable; in robotics, locomotion and manipulation often run  faster than 10 milliseconds, so the same delay can cause physical collapse, hardware damage, or injury. Adversaries can exploit this narrow operational window, for example, LiDAR spoofing (SP-A1) can destabilize motion within a control cycle, before a detector reacts. Security mechanisms that do not operate within these deadlines provide no prevention-they enable only post-incident forensic analysis.
                                    
\input{table}

\input{tableDefenses}

\subsection{A Layer-by-Layer Analysis of Threats and Defenses}

\noindent \textbf{Physical Layer - hardware, power, and safety baseline.}
The physical layer represents the most critical point of potential system compromise. 
Runtime attacks like firmware reflashing (P-A1) can embed persistent malware via debug interfaces, bypassing all software controls. 
Sensor bias (P-A2) distorts physical inputs, inducing silent drift in downstream systems. 
Since these attacks occur below the software stack, conventional detection mechanisms often miss them. This risk is amplified by perimeter-less operation; attackers need not breach an IT facility to reach the asset; they can target the robot in situ (casing, port access, sensor spoofing), then escalate across layers.

\noindent \textbf{Sensing and Perception Layer - Integrity Control Point for Environmental Input.}
The sensing layer functions as an important integrity checkpoint, as sensors must process untrusted environmental signals in real-time. Optical spoofing (SP-A1) injects false LiDAR returns, while resonant ultrasound (SP-A3) destabilizes MEMS gyroscopes. Verifying the authenticity of external inputs in real-time becomes difficult. Defenses like cross-modal verification (SP-D1) and coded illumination (SP-D2) leverage sensor redundancy, but the high volume and rate of sensor data  make exhaustive validation impractical, leaving systems vulnerable to sophisticated physical manipulation techniques.

\noindent \textbf{Data Processing Layer - timing and fusion correctness.}
This layer is governed by strict real-time constraints (typically 1–10,ms). Even minor timing disruptions, such as from cache contention (DP-A1) or biased state estimation (DP-A4), can degrade system stability without triggering alarms. Defenses like timing guards (DP-D1) and memory-safe pipelines (DP-D2) mitigate such risks but often at a latency cost that real-time controllers cannot accommodate.

\noindent \textbf{Middleware Layer - Communication Infrastructure Attack Surface.}
As the backbone of communication, middleware enables high-throughput data exchange. Yet default configurations often favor usability over security. Topic spoofing (MW-A1) exploits missing authentication checks, while replay attacks (MW-A2) exploit missing freshness checks. Although cryptographic protections (MW-D1) and anomaly detection (MW-D2) can address these threats, their deployment is incomplete due to performance considerations, creating exploitable security gaps.

\noindent \textbf{Decision-Making Layer - Autonomous Control Under Adversarial Influence.}
As autonomy shifts toward ML-driven control, threats evolve from conventional faults to adversarial manipulation. Attacks exploit the opacity of AI models: adversarial inputs (DM-A1) mislead perception, while model poisoning (DM-A5) embeds latent backdoors. 
These attacks often avoid triggering system safeguards or error conditions, making detection particularly difficult.
Defensive techniques like adversarial training (DM-D4) and certified inference (DM-D1) show promise but can be computationally costly. 
Compromises in this layer typically result in subtle misbehavior, where robots appear to function normally while executing attacker objectives.

\noindent \textbf{Application Layer - Remote Control Interface and Update Mechanism.}
As the interface for scripts, APIs, and OTA updates, this layer offers the most direct path to remote manipulation. 
Attacks are often semantic: script edits (AP-A1) silently alter behavior without memory corruption; unauthenticated APIs (AP-A2) enable full override. 
The main vulnerability lies in insufficient runtime policy enforcement. While defenses like signed code (AP-D1) and secure OTA mechanisms (AP-D3) exist, they are inconsistently deployed, allowing a breach here to bypass all lower layer protections.

\noindent \textbf{Social Interface Layer - Human-Robot Interaction as Attack Vector.}
This layer manages human-facing interactions through speech, vision, and gesture recognition systems. These modalities present unique attack surfaces, such as inaudible voice commands (SI-A1) or visual impersonation (SI-A3), which can trigger unauthorized actions without user awareness. Defending this inherently open interface is complex. Technical countermeasures like hardware filtering (SI-D1) and robust automated speech recognition (ASR) (SI-D2) help, but attacks often exploit social expectations rather than system flaws. Effective defenses must therefore combine technical safeguards with user trust boundary training (SI-D5) and privacy-first design principles.

%% file: table.tex
\begin{table*}
    \caption{Empirical attack vectors mapped to humanoid architecture.}
    \label{tab:attack_surface}
    \begin{adjustbox}{width=0.999\textwidth,center}
        \begin{tabular}{lp{0.2\textwidth}p{0.97\textwidth}l}
            \toprule
            {\textbf{Attack ID}} & {\textbf{Attack Vector}} & {\textbf{Description}} & {\textbf{Ref.}} \\
            \midrule
            \rowcolor{blue!10} \multicolumn{4}{c}{\textbf{Physical Layer - Hardware Integrity and Energy Manipulation}} \\
            \midrule
            
           P-A1 & Firmware reflashing & Quarta et al. demonstrated that the FlexPendant boot image delivered by the controller lacks code signing, enabling installation of a malicious firmware that persists across reboot; they used this to falsify the User Interface (UI), misleading the operator about robot state. Humanoids mostly use unsigned firmware images and weak update channels. & \cite{Quarta17}  \\

            P-A2 & Sensor bias via local fields & Many studies~\cite{Shoukry15,Cao2023,Shin17} have empirically demonstrated sensor-bias attacks at the modality level (e.g., magnetic/Hall encoders, LiDAR, MEMS IMUs, microphones) and are platform-agnostic for robots that deploy the same sensor type. & \cite{Shoukry15}  \\

             P-A3 & Power rail manipulation & Zhan et al. demonstrated that attackers can manipulate Qi-compliant wireless chargers by modulating their input supply, and control the charging systems to cause overheating and overcharging~\cite{Zhan24}. 
             humanoids with external DC adapters, can force unsafe charging conditions, potentially triggering system resets that disrupt critical control loops. & \cite{Zhan24}  \\

              P-A4 & Field tampering & Belikovetsky et al. weakened a 3D-printed propeller via CAD/print-path manipulation, which later fractured mid-flight and caused a crash; the defect escaped visual inspection \cite{Belikovetsky17}. This research describes how supply chain attacks on suppliers of hardware components of robotics can serve as silent vectors for physical compromise. & \cite{Belikovetsky17}  \\

               P-A5 & Acoustic side-channels & Shah et al. showed that smartphone microphones can fingerprint robot motions and reconstruct workflows from gearbox acoustics, leaking task-level behavior~\cite{Shah22}. LidarPhone used robot vacuum’s LiDAR as a laser mic to eavesdrop on speech via vibrations of nearby objects. humanoids with similar gearboxes/LiDAR, are vulnerable to similar exploitation~\cite{VacuumACPoster}. & \cite{VacuumACPoster,Shah22}  \\

             \midrule
            \rowcolor{green!10} \multicolumn{4}{c}{\textbf{Sensing and Perception Layer - Sensor Fidelity, Privacy, and Signal Integrity}} \\
            \midrule
            
                SP-A1 & Active optical spoofing & Shin et al. demonstrated that injecting timed infrared pulses into VLP-16 LiDAR systems can create phantom detection points or blind sectors~\cite{Shin17}. Cao et al. achieved a 92.7\% success rate in removing approximately 90\% of target obstacles from point clouds, effectively hiding moving pedestrians from detection. These perception-layer attacks pose threat for humanoid robots using similar LiDAR systems. & \cite{Shin17,Cao2023}  \\

                 SP-A2 & Sensor blinding and jamming & Petit et al. experimentally blinded cameras with high-intensity light and spoofed LiDAR with laser pulses, producing false obstacles or loss of returns. On humanoids, it can mislead footstep planning at edges and cause falls~\cite{Petit2015}. & \cite{Petit2015}  \\

                 SP-A3 & Resonant ultrasound and EMI injection & Son et al. destabilized drones by driving MEMS gyroscopes at acoustic resonance~\cite{Son15}; Trippel et al. injected signals into MEMS accelerometers via sound to control their digital outputs~\cite{Trippel17}. On humanoids, similar IMU bias can corrupt state estimation and cascade to balance control, risking falls. & \cite{Son15,Trippel17}  \\

                 SP-A4 & Physical adversarial patches & Thys et al. printed patches that hide persons from YOLOv2, markedly reducing detections in real-world video~\cite{Thys19}; Eykholt et al. crafted sticker attacks that fooled Faster R-CNN in 85.9\% of frames in lab tests (40.2\% outdoor)~\cite{Eykholt18}. On humanoids, such perception-layer evasion can conceal humans/obstacles and propagate errors to planning and navigation. & \cite{Thys19,Eykholt18}  \\

                 SP-A5 & Sensor-based privacy leakage & Giaretta et al. found that Pepper's HTTP interfaces and unauthenticated endpoints expose camera access with plaintext credentials~\cite{Giaretta2018}, Campazas-Vega et al. observed similar unencrypted HTTP-only communication causing data leak~\cite{CampazasVega23}. & \cite{Giaretta2018,CampazasVega23}  \\
                \midrule
            \rowcolor{orange!10} \multicolumn{4}{c}{\textbf{Data Processing Layer - Runtime Pipelines, Memory safety, and Control Loop Consistency}} \\
            \midrule
            
                DP-A1 & Timing channel interference & Li et al. showed that microsecond-scale contention on PREEMPT-RT (real-time Linux) causes deadline misses, demonstrated when a humanoid research platform ROBOTIS OP3's 1 kHz balance loop failed and caused the robot to fall~\cite{Li2024}. Proctor measured similar microsecond-level MCU jitter from cache and bus contention, confirming that timing interference can destabilize torque control loops~\cite{Proctor2001}. & \cite{Li2024,Proctor2001}  \\

                 DP-A2 & Memory safety exploits & CVE-2018-5268 documented how crafted JPEG2000 images trigger heap-based buffer overflows in OpenCV 3.3.1, compromising perception pipelines that process untrusted visual inputs~\cite{CVE20185268}. Alias Robotics identified 86 vulnerabilities in Universal Robots controllers, including buffer overflows in real-time joint control code, demonstrating how data-processing layer attacks can cascade to planning and actuation systems in humanoid robots~\cite{AliasUR2020}. & \cite{CVE20185268,AliasUR2020}  \\

                  DP-A3 & Race conditions and TOCTOU flaws & ROS 2 documented a MultiThreadedExecutor race condition where messages are dequeued before publishers complete their writes, exposing stale or partial data (rclcpp \#1007). This can propagate stale odometry from data-processing to planning and control systems in humanoid robots~\cite{ROS2Race2020}. & \cite{ROS2Race2020}  \\

                   DP-A4 & Estimator bias tampering & Shen et al. showed GPS spoofing can seize multi sensor fusion (MSF) Kalman Filter (KF) localization and induce >10 m of lateral error before fault alarms fired, letting estimator bias cascade into planning and control~\cite{Shen2020}, while Cao et al.  removed obstacles from LiDAR and LiDAR–camera fusion so detectors missed objects in closed-track tests~\cite{Cao2023}. & \cite{Cao2023}  \\

            \midrule
            \rowcolor{violet!10} \multicolumn{4}{c}{\textbf{Middleware Layer - Secure Communications, Naming, and Orchestration}} \\
            \midrule
                MW-A1 & Topic spoofing and participant impersonation & Mayoral et al. showed that spoofed ROS publishers/subscribers let attackers inject commands and seize endpoints, escalating a middleware breach into planning and actuation effects~\cite{Mayoral2020}. Di Luoffo et al. showed credential-masquerading attacks against DDS/SROS2 that allow participant impersonation and message forgery, with unsecured deployments further exposing RTPS discovery identifiers that facilitate spoofing~\cite{diluoffo2019credential}. & \cite{Mayoral2020, diluoffo2019credential} \\

                MW-A2 & Replay and stale frame injection & Lauser et al. identified a replay vulnerability in DDS-Security's encryption protocol that enables stale message injection even with security enabled~\cite{lauser2025ddssecurity}. In humanoids, replayed odometry and IMU data would propagate through localization and balance systems, compromising stability. & ~\cite{lauser2025ddssecurity} \\

                MW-A3 & Plaintext sniffing and data exfiltration & Giaretta et al. demonstrated that SoftBank's Pepper robot transmitted unencrypted RGB camera and audio streams over Wi-Fi, enabling interception of live sensor data and exposure of personally identifiable information (PII)~\cite{Giaretta2018}. & \cite{Giaretta2018} \\

                MW-A4 & Discovery abuse and denial-of-service (DoS) & Cibrario et al. demonstrated SlowROS, a low rate DoS attack against ROS Master that exploits missing connection-closure timeouts and causes system-wide ROS disruptions. Humanoids rely on ROS middleware for control topics such as /joint\_angles and /imu, a comparable flood could stall message delivery and destabilize balance controllers~\cite{CibrarioBertolotti2025}. & \cite{CibrarioBertolotti2025} \\
            
            \midrule
            \rowcolor{red!10} \multicolumn{4}{c}{\textbf{Decision-Making Layer - Compromise of Autonomy and Control Logic}} \\
            \midrule
                DM-A1 & Adversarial Examples in Perception and Policy & Chen et al. showed that a 5 cm adversarial patch inflated visual SLAM error by 7.3× and caused RGB-D tracking loss within 20 frames~\cite{Chen2024AoR}. Buddareddygari et al. demonstrated that adversarial roadside textures misled deep RL driving policies, inducing emergency stops or wrong-way routing within seconds~\cite{Buddareddygari2022RLSign}. In humanoids, such patches could misclassify gestures or inject phantom obstacles, undermining decision making. & \cite{Chen2024AoR, Buddareddygari2022RLSign} \\

               DM-A2 & Sequential perturbation of control policies & Shi et al. demonstrated that injecting low-magnitude, time-correlated perturbations into IMU readings destabilized the ANYmal C quadruped, causing falls in 87\% of trials despite model predictive control fallbacks. The attack gradually corrupted state estimates across sequential control cycles~\cite{Shi2024ANYmal}. Humanoids fusing joint proprioception with inertial data face similar gradual destabilization risks. & \cite{Shi2024ANYmal} \\

            \bottomrule
        \end{tabular}
    \end{adjustbox}

\end{table*}

\addtocounter{table}{-1}

\begin{table*}
    \caption{Empirical attack vectors mapped to humanoid architecture.}
    \label{tab:attack_surface}
    \begin{adjustbox}{width=0.999\textwidth,center}
        \begin{tabular}{lp{0.2\textwidth}p{0.95\textwidth}l}
            \toprule
            {\textbf{Attack ID}} & {\textbf{Attack Vector}} & {\textbf{Description}} & {\textbf{Ref.}} \\
            \midrule               
                DM-A3 & Concept drift and sensor fusion spoofing & Shen et al. introduced FusionRipper, a GPS spoofing attack that stealthily biases multi-sensor fusion localization without triggering health alarms. On Baidu's Apollo, it laterally shifted the MSF output by over 10 m within minutes while reporting OK status. Humanoids using visual-inertial odometry, ultra wide band (UWB) beacons, or floor-marker fusion may similarly be misled via staged bias injections that exploit KF latency. & \cite{Shen2020} \\
                DM-A4 & Reward hacking and specification gaming & Krakovna et al. catalogued specification-gaming cases in reinforcement learning, such as CoastRunners agents repeatedly hitting bonus targets instead of finishing races. Domestic humanoid rewarded for staying active near its charging dock might simply spin in place to maximize reward, rather than performing its intended household duties~\cite{Krakovna2020SpecGaming}. & \cite{Krakovna2020SpecGaming} \\

                DM-A5 & Model poisoning and physical backdoors & Wang et al. proposed TrojanRobot, a backdoor attack that embeds visual triggers into vision-language model (VLM) manipulation policies. Testing on a UR3e robotic arm showed that mundane objects stealthily alter action outputs while maintaining normal performance otherwise~\cite{wang2024trojanrobot}. Humanoids using similar vision-language-action pipelines for grasping or tool use face comparable risks of stealthy policy hijacking. & \cite{wang2024trojanrobot} \\

                DM-A6 & Finite State Machine (FSM) corruption and logic bombs & Govil et al. introduced ladder logic bombs (LLBs)-malicious code embedded in PLC programs that remains dormant until triggered, then overwrites control states or sensor outputs while evading detection~\cite{Govil2017LLB}. In humanoids, similar latent attacks could hide within gait or balance controllers, activating during specific pose transitions or sensor events. & \cite{Govil2017LLB} \\

                DM-A7 & Control parameter tampering & Trend Micro's Rogue Robots study demonstrated that subtle tampering with ABB IRB 140 control parameters-such as servo gains or sensor offsets-introduces undetected motion deviations that compromise safety~\cite{TrendMicro2017RogueRobots}. In humanoids, manipulation of joint limits or kinematic parameters could degrade motion safety without triggering controller alarms. & \cite{TrendMicro2017RogueRobots} \\

                DM-A8 & Planning logic manipulation & Vemprala et al. demonstrated that adversaries can manipulate optimization-based motion planners to produce unsafe trajectories by perturbing cost functions~\cite{Vemprala2021PlannerAttack}. Since humanoids rely on cost-map planning, similar tampering through false sensor inputs or corrupted avoidance rules could misdirect navigation and compromise safety. & \cite{Vemprala2021PlannerAttack} \\

                DM-A9 & Malicious OTA updates & NHTSA classified firmware OTA updates as high-impact threat vector, since compromised OTA servers can silently corrupt entire system before detection. Humanoids face similar risks of stealthy sabotage that bypasses safety checks~\cite{NHTSA2022BestPractices}.& \cite{NHTSA2022BestPractices} \\

                DM-A10 & Federated learning poisoning & Bagdasaryan et al. demonstrated that a single malicious participant in federated learning can replace the global model with a backdoored version, preserving accuracy while achieving 100\% success on hidden malicious objectives~\cite{Bagdasaryan2020BackdoorFL}. This vulnerability could enable stealthy manipulation of decision-making policies across multiple deployed humanoid units. & \cite{Bagdasaryan2020BackdoorFL} \\

                DM-A11 & Trojaned pretrained models and supply chain threats & Wang et al. embedded a backdoor into a pretrained VLM controlling a UR3e robot arm. The model operated normally until visual triggers activated attacker-specified manipulation routines, demonstrating that trojaned models can survive deployment undetected~\cite{wang2024trojanrobot}. Humanoid systems using third-party or cloud-hosted models face similar supply-chain risks of malicious behaviors that activate only under specific conditions. & \cite{wang2024trojanrobot} \\
                
            \midrule
            \rowcolor{yellow!10} \multicolumn{4}{c}{\textbf{Application Layer - Task Scripts, APIs, and OTA Operations}} \\
            \midrule
                AP-A1 & Script and state-machine corruption & 
                Quarta et al. demonstrated that modifying ABB RAPID control code silently alters robot behavior while status displays showed normal operation~\cite{Quarta17}. Pogliani et al. showed that URControl robot's network interface accepted malicious URScript commands, enabling remote code execution with root privileges. In humanoids, such attacks could modify the behavior trees to alter movement patterns or disable safety checks~\cite{Pogliani2019ConnectedFactory}. & \cite{Quarta17,Pogliani2019ConnectedFactory}\\

                AP-A2 & Unauthenticated APIs and remote override & DeMarinis et al. found over 100 ROS systems exposed on the public internet, enabling unauthenticated remote control of robots through basic commands like /cmd\_vel~\cite{DeMarinis2019ROSScan}. Giaretta et al. discovered that Pepper's NAOqi API lacked authentication, allowing attackers to access cameras, microphones, and disable safety functions~\cite{Giaretta2018}.  & \cite{DeMarinis2019ROSScan,Giaretta2018} \\

                AP-A3 & CI/CD supply chain compromise & Oyeronke analyzed firmware supply chains in CI/CD pipelines and showed that attackers could tamper with dependencies, inject code during builds, or steal signing keys to distribute malicious images as legitimate OTA updates~\cite{firmwarecicd}. In humanoids, such a poisoned pipeline could enable the silent deployment of trojaned firmware across fleets of robots.  & \cite{firmwarecicd} \\

                AP-A4 & Sim-to-real discrepancy and digital-twin overload & Xu et al. showed that a single malicious node can saturate cloud multi-robot services, causing 35\% packet loss and reducing robot velocity to 0.04-0.05 m/s, leading to mission failures and crashes~\cite{Xu2021}. This overload disrupts digital twin synchronization: delayed updates create twin-physical misalignment, causing plans to be computed on stale state data~\cite{Tandigital}. Since latency drives the sim-to-real gap, these synchronization delays result in degraded or unsafe behaviors when control policies expecting timely feedback execute on outdated measurements~\cite{Tan-sim}. & \cite{Xu2021,Tan-sim,Tandigital}  \\

                AP-A5 & Runtime parameter tampering & Quarta et al. demonstrated that modifying controller parameters degraded accuracy and induced instability in ABB IRB140 robot, with configuration changes propagating from application to servo levels~\cite{Quarta17}. For humanoids, similar attacks on ROS-hosted gait parameters (YAML/URDF) could silently destabilize walking while bypassing high-level safety checks. & \cite{Quarta17} \\

            \midrule
            \rowcolor{cyan!10} \multicolumn{4}{c}{\textbf{Social Interface Layer - Speech, Gesture and the Human‑Trust Channel}} \\
            \midrule    
                SI-A1 & Inaudible and line-of-sight command injection & Zhang et al. showed that ultrasound-modulated, inaudible commands can activate and control voice assistants at distances~\cite{Zhang2017DolphinAttack}. Sugawara et al. showed that amplitude-modulated lasers inject commands through closed glass~\cite{Sugawara2020LightCommands}. This enables silent takeover of social interfaces in humanoids that can cascade to decision and application layers. & \cite{Zhang2017DolphinAttack, Sugawara2020LightCommands} \\

                SI-A2 & Synthetic or hidden audio & Yuan et al. embedded automatic speech recognition (ASR) commands into songs~\cite{Yuan2018CommanderSong}. Li et al. crafted adversarial music that hides ghost commands which robot dialogue systems’ ASR models execute with high success rates while sounding normal to humans~\cite{Liu2024PhantomOpera}. Humanoids working on voice commands can be exploited through social interfaces with effects cascading to decision/application layers. & \cite{Yuan2018CommanderSong,Liu2024PhantomOpera} \\

                SI-A3 & Visual spoofing and adversarial cues & Thys et al. demonstrated that printed adversarial patches reduced YOLOv2 person detection allowing 61-74\% of people to bypass detection~\cite{Thys19}. Wu et al. used projected structured-light patterns from 2D photos to spoof commercial 3D face authentication with 79.4\% success (depth-only)~\cite{Wu2023DepthFake}. For humanoids, adversarial clothing can enable user impersonation. & \cite{Thys19,Wu2023DepthFake}  \\

                SI-A4 & Covert A/V eaves-dropping via social channels &  Giaretta et al. found that Pepper's web interface exposed credentials, enabling privilege escalation for covert surveillance~\cite{Giaretta2018}. IOActive demonstrated remote access to cameras and microphones on Pepper/NAO and UBTECH Alpha-2 robots~\cite{oruma}. Denning et al. showed that consumer telepresence robots could be compromised to stream live video for home surveillance~\cite{denning2009spotlight}. Such compromises in humanoids enables covert surveillance and intrusion of privacy. & \cite{Giaretta2018,denning2009spotlight,oruma}  \\

                SI-A5 & Trust manipulation and social engineering & Aroyo et al. used iCub robot in social-engineering scenarios where participants disclosed sensitive data and were induced to gamble their winnings, demonstrating overtrust-driven risk-taking. Wolfert et al. showed Pepper could tailgate into secure areas and convinced all participants to insert unknown USB devices. Such compromise can enable phishing attacks and coercion of unsafe user actions. & \cite{Belpaeme2019RobotPersuasion,Aroyo2018TrustSE}  \\           
            \bottomrule
        \end{tabular}
    \end{adjustbox}

\end{table*}

%% file: tableDefenses.tex
\begin{table*}
    \caption{Defense mechanisms across humanoid architecture layers.}
    \label{tab:defences}
    \begin{adjustbox}{width=0.999\textwidth,center}
        \begin{tabular}{lp{0.2\textwidth}p{0.95\textwidth}l}
            \toprule
            {\textbf{Defense ID}} & {\textbf{Defense Vector}} & {\textbf{Description}} & {\textbf{Ref.}} \\
            \midrule
            \rowcolor{blue!10} \multicolumn{4}{c}{\textbf{Physical layer - Hardware Integrity and Energy Manipulation}} \\
            \midrule
            
           P-D1 & Secure boot & Siddiqui et al. implement hardware root-of-trust for Xilinx FPGAs with authenticated boot bitstreams~\cite{FPGAsecureBoot20}, while NVIDIA Jetson uses on-die BootROM with PKC fuses for signed boot images~\cite{NVIDIA_L4T_SecureBoot,NVIDIA_Jetson_SecureBoot_r35_4_1}. This protects humanoid robots using SPI flash boot and FPGA motor controllers from reflashing attacks and privilege escalation. & \cite{NVIDIA_L4T_SecureBoot,NVIDIA_Jetson_SecureBoot_r35_4_1,FPGAsecureBoot20}  \\

            P-D2 & Redundant encoders & Campeau et al. fused optical encoders with Hall-effect sensors using Kalman filtering on a 2-DoF assistive robot, detecting sensor faults through residual analysis~\cite{CampeauLecours17}. This applies to humanoids' balance-critical joints like knees and ankles. & \cite{CampeauLecours17}  \\

             P-D3 & Rail-level cutouts & Ismail et al. described battery-disconnect units with contactor and fast-acting fuses that isolate DC buses during over-current events~\cite{Ismail2024}. For humanoids with 48-800V actuator rails, these cutouts prevent power faults from cascading into compute/control systems. & \cite{Ismail2024}  \\

             P-D4 & Tamper evidence & Vidakovic classified conductive break-trace meshes as tamper-evident, providing persistent evidence of tampering that can trigger detection responses. Paley et al. used active PCB meshes measuring trace resistance to detect drilling/cuts and trigger shutdown~\cite{PaleyHB16}. In humanoids, this enables tamper evidence and safe-mode responses against physical attacks. & \cite{Vidakovic23,PaleyHB16}  \\

             P-D5 & Leak-resilient mechanics & Matlack et al. demonstrated broadband vibration absorption using composite meta-structures, achieving vibration and gear-whine reductions with damping treatments~\cite{Matlack15}. These retrofittable CLD/meta-material patches apply to humanoid harmonic-drive enclosures to suppress acoustic side-channels. & \cite{Matlack15}  \\

            \midrule
            \rowcolor{green!10} \multicolumn{4}{c}{\textbf{Sensing and Perception Layer - Sensor Fidelity, Privacy, and Signal Integrity}} \\
            \midrule
            
                SP-D1 & Cross-modal verification and temporal coherence & You et al. introduced temporal-consistency checks across consecutive LiDAR frames, detecting spoofed objects with >98\% accuracy at 41 Hz. Hau et al.'s Shadow-Catcher validates detections by verifying physically consistent 3D shadows, achieving >94\% accuracy in 3-21 ms. In humanoids with multiple depth sensors it can prevent spoofed LiDAR data. & \cite{You2021,Hau2021}  \\

                 SP-D2 & Coded illumination and pulse signatures & Wang et al. introduced pseudorandom-modulation quantum-secured LiDAR, using randomized photon properties and code correlation for ranging against spoofing~\cite{wang2015}. Yu et al. developed true-random coded photon-counting LiDAR to generate unique pulse codes~\cite{Yu2019}. Integrating these into humanoids' active depth sensors could enable real-time anti-spoofing. & \cite{wang2015,Yu2019}  \\

                 SP-D3 & Acoustic and EMI hardening & Jeong et al. developed a denoising autoencoder that recovered IMU signals under acoustic interference against acoustic injection attack~\cite{Jeong2023}. Kune et al. evaluated shielding, differential inputs, and adaptive filtering to attenuate EMI-borne artifacts~\cite{Kune13}. In humanoids they can stabilize IMU-based balance and prevent bias from propagating to gait control.& \cite{Jeong2017,Kune13}  \\

                 SP-D4 & Robust perception via adversarial training & Liu et al.'s ARMRO detected ~95\% of adversarial patches in vision transformers with ~1\% accuracy loss~\cite{Liu2023}. Hofman et al.'s X-Detect flags digital/physical patches in real-time with low false positives~\cite{hofman_x-detect_2024}. These defenses enable humanoids to mask patched regions before visual tracking, SLAM, and object recognition. & \cite{Liu2023}  \\

                 SP-D5 & On-device encryption and data hygiene & Kim et al. demonstrated encryption and authentication of low-rate sensor streams~\cite{Kim2020}. Mi et al. demonstrated accurate face identification despite stripping all information other than required for face identification~\cite{Mi2024}. In humanoids, these can be used to streams from low rate sensors (IMU, joint states) and preserve privacy. & \cite{Kim2020,Mi2024}  \\

            \midrule
            \rowcolor{orange!10} \multicolumn{4}{c}{\textbf{Data Processing Layer - Runtime Pipelines, Memory Safety, and Control Loop Consistency}} \\
            \midrule
            
                DP-D1 & Timing guards & Abaza et al. implemented Logical Execution Time (LET) for bounded and scheduled execution of periodic task (e.g., IMU-estimator-controller-actuators) with  
                negligible end-to-end jitter~\cite{Abaza2024}. Ye et al. used LET windows and deadline guards to keep end-to-end timing predictable and prevent late, unstable commands from reaching the actuators~\cite{Ye2023}. For humanoids, these guards bounded timing and suppressed tardy outputs before actuation, stabilizing balance and gait. & \cite{Ye2023,Abaza2024}\\

                 DP-D2 & Memory safety & Schmidt et al. rewrote NAO-class Extended KF/vision in Rust, eliminating common memory-safety bugs for ≈6\% CPU overhead~\cite{HULKs2023}; Deng et al. evaluated address space layout randomization and stack canaries to contain residual C/C++ faults~\cite{Deng2022}. For humanoids, porting balance/perception nodes to Rust and hardening legacy nodes reduces crash risk. & \cite{HULKs2023,Deng2022}  \\

                 DP-D3 & Race-proof queues & Paccagnella et al. exposed deletable audit events in kernel buffers before commit~\cite{Paccagnella2020}. Ahmad et al.'s HARDLOG sealed events synchronously with tamper-proof devices, eliminating pre-commit races~\cite{hardlog_ahmad}. Wagner et al. used TPM quotes for DDS peer attestation~\cite{Wagner2024}. For humanoids, protected audit devices and attested DDS membership can prevent tampering. & \cite{Wagner2024,hardlog_ahmad,Paccagnella2020}  \\

                DP-D4 & Parameter attestation and sanity filters & 
                Kuang et al.'s DO-RA attests runtime EKF parameters via MAC checks each cycle, triggering failsafe on mismatch~\cite{kuang2020}. Tanil et al. use EKF innovation-$\chi^2$ gating for sensor exclusion~\cite{tanil2018}, and Bloesch et al. cross-validate exteroceptive pose against IMU/kinematics~\cite{Bloesch-RSS-12}. This layered approach detects parameter tampering and prevents sensor spoofing. & \cite{kuang2020,tanil2018,Bloesch-RSS-12}  \\

            \midrule
            \rowcolor{violet!10} \multicolumn{4}{c}{\textbf{Middleware Layer - Secure Communications, Naming, and Orchestration}} \\
            \midrule

                MW-D1 & Cryptographic hardening (DDS-Security / SROS2) & SROS2 provided mutual authentication, AES-GCM transport protection, and XML-based authorization policies that reject unauthorized publishers/subscribers on control topics of robot~\cite{Deng2022}. For humanoids, enforcing SROS2 security features with automated X.509 provisioning constrain remote takeover of motion-critical topics. & \cite{Deng2022} \\

                MW-D2 & Intrusion detection and policy enforcement & Rivera et al.'s ROSDN enforced policies on each topic policies via SDN firewall on Turtlebot3~\cite{Rivera2019}. Soriano et al.'s RIPS detected unauthorized subscriptions and commands on TIAGo using rule-based filtering, triggering mitigations limiting video leakage~\cite{SorianoSalvador2024}. For humanoids, these policy gates on DDS topics prevents command spoofing and data exfiltration. & \cite{Rivera2019,SorianoSalvador2024}  \\

                MW-D3 & Information-flow control & Pandya et al.'s Picaros enforced decentralized data flow control in ROS 2 using secrecy labels. For humanoids, labeling robot topics confines downstream use and prevents unlabeled nodes from exfiltrating data across distributed deployments.~\cite{Picaros2024}. & \cite{Picaros2024}  \\

                MW-D4 & QoS shaping and freshness enforcement & Abaza et al. maintained negligible ROS 2 jitter using LET and reservation servers~\cite{Abaza2024}, while DDS Lifespan/Deadline policies bound stale deliveries. Wagner et al. used nonce-fresh attestation for measured peer admission~\cite{Wagner2024}. This combination prevents stale commands from reaching humanoid actuators. & \cite{Wagner2024,Abaza2024}  \\
                \midrule
 
            \rowcolor{red!10} \multicolumn{4}{c}{\textbf{Decision‑Making Layer - Compromise of Autonomy and Control Logic}} \\
            \midrule

                DM-D1 & Redundant and certified decision-making & Gandhi et al.'s proposed bounded-time interaction for multi-robot systems, maintaining mission progress despite faulty/malicious robots with modest overhead~\cite{Gandhi2025RoboRebound}. Lutjens et al. computed certified action selection under bounded observation perturbations, rejecting actions violating worst-case safety margins at runtime. For humanoids, fleet redundancy and certified action filters prevent compromised modules or noisy sensors from cascading into unsafe behavior.~\cite{Lutjens2020CertifiedRL}. & \cite{Lutjens2020CertifiedRL,Gandhi2025RoboRebound}  \\

                DM-D2 & Adversary-aware planning & Shi et al. formulated RMOP that maintains bounded team rewards under $\alpha$ robot dropouts with constant-factor guarantees and online MCTS variants. Soriano et al. built ROS 2 intrusion prevention with topic monitoring and rule-based System Modes mitigations on TIAGo~\cite{SorianoSalvador2024}. For humanoids, $\alpha$-tolerant planning and IDS-driven mode switches preserve mission progress and enable safe idle during abnormal communications. & \cite{Shi2023RMOP,SorianoSalvador2024}  \\

            \bottomrule
        \end{tabular}
    \end{adjustbox}
\end{table*}

\addtocounter{table}{-1}

\begin{table*}
    \caption{Defense mechanisms across humanoid architecture layers.}
    \label{tab:defences}
    \begin{adjustbox}{width=0.999\textwidth,center}
        \begin{tabular}{lp{0.2\textwidth}p{0.95\textwidth}l}
            \toprule
            {\textbf{Defense ID}} & {\textbf{Defense Vector}} & {\textbf{Description}} & {\textbf{Ref.}} \\
            \midrule

                DM-D3 & Simulated adversary modules (shadow probe) & Shi et al. trained a small disturbance policy and ran it alongside the controller; the probe predicted when tiny IMU/force perturbations would make a candidate action unsafe, after which a safety filter discarded the action and logged a reproducible test case~\cite{Shi2024ANYmal}. Pinto et al. cast such disturbance agents as adversaries in robust RL~\cite{pinto2017}. In humanoids, robustness can be improved without retraining by evaluating each step/torque command in parallel and blocking those that failed a worst-case check. & \cite{Shi2024ANYmal,pinto2017} 
                \\
                
                DM-D4 & Adversarial training and risk-aware policies & Xie et al. adversarially trained visual-motor policies with learned visual perturbers for better generalization~\cite{Xie2025DualAgent}, while Han et al. co-trained with embodied disturbance agents to improve locomotion robustness~\cite{Han2024Agility}. Lutjens et al. enforced certified safety filters rejecting unsafe actions under bounded observation noise~\cite{Lutjens2020CertifiedRL}. For humanoids, combining adversarial training with runtime certified filters hardens control against lighting shifts and sensor errors without retraining. & \cite{Xie2025DualAgent, Lutjens2020CertifiedRL,Han2024Agility}   \\

                DM-D5 & Poisoning-resilient learning & Liu et al. trained behavior-cloning networks with median-of-means loss by splitting demonstrations into groups, averaging each group, then optimizing toward the median average. This down-weights poisoned/low-quality demonstrations while maintaining clean-data performance, directly applicable to humanoids learning from noisy or attackable teleop logs~\cite{Liu2022RobustIL}. & \cite{Liu2022RobustIL}  \\

                DM-D6 & Backdoor detection & Fan proposed PeerGuard that enables agents to cross-check peers and generate explicit reasoning for inconsistencies, detecting poisoned agents with high accuracy in LLM multi-agent evaluations~\cite{Fan2025PeerGuard}. For humanoids, a peer safety agent could validate action rationales against sensor facts and task intent before execution, flagging compromised policies at runtime. & \cite{Fan2025PeerGuard}  \\

                DM-D7 & Secure model updates and storage & Plappert and Fuchs developed TPM-anchored OTA scheme that authenticated update bundles and derived per-update keys only when TPM policies validated, with all symmetric keys kept inside the TPM. They reported kB-scale metadata overhead with TPM enforcement~\cite{Plappert2023SecureOTA}. For humanoids, signing firmware/ML models, sealing keys to measured state, and using Linux integrity measurement architecture (IMA)/ extended verification module (EVM) or fs-verity prevents forged/replayed packages and blocks unsigned model execution.  & \cite{Plappert2023SecureOTA}  \\

            \midrule
            \rowcolor{yellow!10} \multicolumn{4}{c}{\textbf{Application Layer - Task Scripts, APIs, and OTA Operations}} \\
            \midrule

                AP-D1 & Signed scripts and runtime monitors & Colledanchise et al. generated runtime monitors from behavior tree specifications that detected requirement violations on the IIT R1 robot during service tasks. Linux IMA/EVM or fs-verity enforced BT/YAML integrity with userspace signing, preventing silent edits from loading~\cite{Colledanchise2021}. For humanoids, this can block tampered task scripts execution and raise faults before unsafe transitions reach actuators. & \cite{Colledanchise2021}  \\

                AP-D2 & API gateways and intrusion detection & ROS 2 enforced per-node permissions with DDS Security's signed XML (SROS2), restricting topic access rights. Rivera et al. filtered per-topic/GUID flows using an SDN firewall (ROSDN) on Turtlebot3 and enabled policy-based blocking of forged traffic~\cite{Rivera2019}. Soriano et al. built RIPS for ROS 2 with a rule DSL (topic matching, mitigations via System Modes) and demonstrated live prevention on a social robot~\cite{SorianoSalvador2024}. For humanoids, strict SROS2 permissions and policy gates on motion-critical topics contain spoofing/exfiltration before affecting balance and gait.& \cite{Rivera2019,SorianoSalvador2024}  \\

                AP-D3 & Secure OTA and reproducible CI & Plappert et al. sealed update-signing keys in TPM with bootloader-enforced signature/version validation, causing only kB-scale metadata overhead~\cite{Plappert2023SecureOTA}. Combined with signed continuous integration (CI) receipts (supply chain levels for software artifacts) and device file checks (fs-verity/IMA), this ensures firmware/ML models match CI builds exactly, preventing forged or replayed packages. This applies to humanoids updating firmware and policies over ROS 2/DDS. & \cite{Plappert2023SecureOTA}  \\

                AP-D4 & Twin hardening and domain randomization & Xu et al. showed that limiting  per-robot API calls can prevent cloud-planner latency inflation~\cite{Xu2021}, Muratore et al. implemented domain randomization to train policies that tolerate sim-to-real mismatch~\cite{Muratore2022}, and Betzer et al.performed runtime checks against a digital twin and triggered a safe stop when live traces violated specified properties~\cite{Betzer2024}. For humanoids, rate-limiting twin APIs to prevent planning stalls, train with domain randomization for sim-to-real robustness, and twin-based runtime verification narrows the reality gap and halts unsafe divergence early. & \cite{Xu2021,Muratore2022,Betzer2024}  \\

                AP-D5 & Parameter whitelists and dynamic guards & Quarta et al. showed that controller parameter changes on ABB IRB140 that caused accuracy/safety violations without alarms can be prevented by locking safety-critical gains as immutable/signed files (IMA/EVM or fs-verity), restricting parameter access to authorized nodes (SROS2), and adding runtime guards that compare in-memory values to signed baselines with fault trips on divergence~\cite{Quarta17}. For humanoids, sealing EKF/controller gains and guarding edits blocks instability created from parameter tampering. & \cite{Quarta17} \\

            \midrule
            \rowcolor{cyan!10} \multicolumn{4}{c}{\textbf{Social Interface Layer - Speech, Gesture, and the Human‑Trust Channel}} \\
            \midrule    

                SI-D1 & Hardware filtering and signal checks & Zhang et al. (EarArray) and Liu et al. (MicGuard) detected inaudible/optical command injections at the microphone front-end before ASR, blocking non-human voice inputs on the human-robot interaction (HRI) channel~\cite{Zhang2021EarArray,Liu2024MicGuard}. For humanoids with always-on voice user interfaces, these SI-layer filters prevent malicious spoken commands from reaching middleware/control, containing cross-layer escalation.& \cite{Zhang2021EarArray,Liu2024MicGuard}  \\

                SI-D2 & Robust ASR and speaker authentication & He et al. built an active ultrasonic canceller that transmits guard tones, causing mic non-linearity to fold inaudible commands into audible bands where adaptive filtering detects and subtracts injected content~\cite{CommandShield2022}. For humanoids with always-on voice interfaces, mounting head emitters and running cancellers before ASR blocks inaudible triggers and prevents cross-layer escalation.. & \cite{CommandShield2022}  \\

                SI-D3 & Anti-spoof vision & Liu et al. showed that flashing a short, random light-CAPTCHA and checking the RGB reflections verifies a live face with a standard camera~\cite{Liu2019AuroraGuard}. Wu et al. (DepthFake) harden depth-based liveness by randomizing the projector pattern and enforcing 3D-consistency checks~\cite{Wu2023DepthFake}. For humanoids, combining light-CAPTCHA on the RGB path with depth-sensor liveness stops video spoofs at the social-interface layer, preventing credential abuse from propagating to control. & \cite{Liu2019AuroraGuard,Wu2023DepthFake}  \\

                SI-D4 & Sensor access control and privacy by design & Soriano et al.'s RIPS-ROS2 enforced per-topic rules on social-interface endpoints (e.g., /asr/result, /teleop/*) and triggered safe modes, blocking unauthorized commands with ≈0.8-1.2s reactions on TIAGo robot~\cite{SorianoSalvador2024}. Martin et al. added on-device redaction masking biometric identifiers in HRI streams before off-robot use~\cite{martin2025towards}. For humanoids, pairing RIPS-style SI gates with edge redaction can prevent command abuse and PII leakage beyond the SI layer. & \cite{SorianoSalvador2024, martin2025towards}  \\

                SI-D5 & Script integrity and trust boundary training & Colledanchise et al. generated monitors from signed Behavior-Tree specs that abort illegal transitions in <4ms on IIT R1 robot, stopping tampered task scripts before reaching actuators~\cite{Colledanchise2021}. Saunderson et al. showed verbal warnings before risky requests cut user compliance, reducing unsafe intent confirmation at voice interfaces~\cite{Saunderson2021Persuasion}. For humanoids, signing BT/YAML with runtime monitoring and requiring SI safety preambles for high-risk intents prevents bad commands from propagating beyond the social interface. & \cite{Saunderson2021Persuasion,Colledanchise2021}  \\
                
            \bottomrule
        \end{tabular}
    \end{adjustbox}
\end{table*}

%% file: gap_analysis.tex
\section{RISK-MAP: A Quantitative Method for Humanoid Security Assessment}
\label{sec:risk_map}

In Section~\ref{sec:attack_surface}, we presented a qualitative overview of the threats and defenses relevant to humanoid platforms. To move from description to prioritization, we introduce RISK-MAP (Robot Integrated Security Key-metric Mapping \& Assessment Platform), a quantitative method that systematically assesses a humanoid’s security posture. This section defines the method and demonstrates it on three humanoids.

\subsection{The RISK-MAP Method}
\label{subsec:risk_map_method}

The RISK-MAP method is built on the set of 39 attack vectors $\mathcal{A} = \{a_1, \ldots, a_{39}\}$ and 35 defense mechanisms $\mathcal{D} = \{d_1, \ldots, d_{35}\}$ identified in our SoK. The method comprises four steps: first establishing baseline threat severity across all identified attacks, then adjusting these assessments for platform-specific applicability, modeling the effectiveness of available defenses for the target platform, and finally aggregating these components into a comprehensive security score.

\noindent\textbf{Step 1: Baseline Threat Severity Assessment.}
To prioritize threats, we first calculate a generic \textbf{Severity} score ($\omega_i$) for each of the 39 attack vector $a_i$. This score is the product of the attack's likelihood and its potential impact.
\begin{equation}
  \text{Severity } \omega_i = \lambda_i \times \iota_i,\quad \text{where } i \in \{1, \ldots, 39\}
\end{equation}
The severity score combines two components:
\begin{itemize}[left=0pt, nosep, leftmargin=*]
    \item A \textbf{likelihood score} ($\lambda_i \in [0,1]$) representing the probability of occurrence, derived from historical data (e.g., CVEs), threat intelligence reports, and expert feasibility assessment.
    \item An \textbf{impact score} ($\iota_i \in [0,1]$) quantifying  potential consequences of a successful attack, based on the following scale:
    \begin{itemize}[leftmargin=*, topsep=0pt, itemsep=0pt]
         \item \textbf{0.0:} Negligible impact and consequences
        \item \textbf{0.2:} Minor disruption; brief degradation
        \item \textbf{0.4:} Functional loss; data issues
        \item \textbf{0.6:} Major failure; prolonged outage
        \item \textbf{0.8:} Safety-critical; risk to people/property
        \item \textbf{1.0:} Catastrophic failure; severe safety consequences
    \end{itemize}
\end{itemize}
This produces a generic severity vector $\Omega = \{\omega_1, \ldots, \omega_{39}\}$, which prioritizes threats independently of a specific platform.

\noindent\textbf{Step 2: Platform-Specific Adjustments:}
Not every attack vector applies to every humanoid platform. To take this into account, we define a binary \textbf{Applicability Vector} $Z^P$, where $Z_i^P=1$ if attack $a_i$ is relevant to platform $P$, and $Z_i^P=0$ otherwise. We then adjust the severity score for the specific platform by zeroing the weight of inapplicable attacks:
\[
\tilde{\omega}_i^{P} = Z_i^{P} \times \omega_{i}
\]
This creates the platform-adjusted severity vector, $\tilde{\Omega}^P$, which ensures that inapplicable attacks are not reflected in the final score.

\noindent\textbf{Step 3: Modeling Defense Effectiveness:}
We evaluate defenses in two stages: assessing their baseline capability and their platform-specific implementation.

\noindent First, we construct the baseline \textbf{Coverage Matrix} $\Gamma \in \mathbb{R}^{39 \times 35}$. Each entry $\gamma_{ij}$ represents the capability of defense $d_j$ to mitigate attack $a_i$, using the following scale. These values are guided by empirical evidence of effectiveness in the literature.
\begin{itemize}[left=0pt, nosep, leftmargin=*]
    \item \textbf{0.00: No mitigation capability.} The defense has no effect on the attack.
    \item \textbf{0.25: Limited (25\% effectiveness).} The defense provides minimal, often indirect, mitigation.
    \item \textbf{0.50: Moderate (50\% effectiveness).} The defense significantly impedes the attack but can be bypassed.
    \item \textbf{0.75: Strong (75\% effectiveness).} The defense reliably mitigates the attack under most conditions.
    \item \textbf{1.00: Complete (100\% effectiveness).} The defense has been proven to entirely block the attack's core mechanism. 
\end{itemize}

\noindent Second,We define an \emph{Implementation Vector} $\mu^P$ for each platform $P$ that measures how well each defense $d_J$
is deployed. The implementation level is scored as:

\begin{itemize}[left=0pt, nosep, leftmargin=*]
    \item \textbf{0.0:} Defense completely absent.
    \item \textbf{0.25:} Minimal implementation; with major gaps.
    \item \textbf{0.50:} Partial implementation; moderately effective.
    \item \textbf{0.75:} Near-complete implementation; minor gaps.
    \item \textbf{1.0:} Fully implemented; best practices met.
\end{itemize}
The \emph{Effective Coverage } ($\varepsilon_{ij}^P$) of defense $d_j$ against attack $a_i$ on platform $P$ is then the product of its baseline capability ( $\gamma_{ij}$) and its specific implementation ($ \mu_j^P$): 
\[
\text{Effective Coverage }\varepsilon_{ij}^P = \gamma_{ij} \times \mu_j^P
\]

\noindent\textbf{Step 4: Calculating Final Scores:}
To compute the combined effectiveness of multiple defenses against a single attack $a_i$, we assume that the defenses act independently. The \emph{total coverage} for that attack on platform $P$ ($\kappa_i^P$) is:
\[
\text{Total Coverage }\kappa_i^P = 1 - \prod_{j=1}^{35} \left(1 - \varepsilon_{ij}^P\right)
\]
The final, aggregated \textbf{RISK-MAP score} for platform $P$ is the severity-weighted average of the coverage of all defenses against an attack, presented as a percentage:
\[
\text{RISK-MAP}_{\%}^{P} =
\left(
\frac{
\displaystyle \sum_{i=1}^{39} \tilde{\omega}_i^{P} \kappa_i^{P}}
{\displaystyle \sum_{i=1}^{39} \tilde{\omega}_i^{P}}
\right) \times 100.
\]

\noindent \textbf{Layer-Specific Diagnostic Analysis:}
While the aggregated score is useful for high-level comparison, it can mask critical architectural weaknesses. To identify such localized weaknesses, we also compute a \textbf{LayerScore} for each of the seven architectural layers ($\ell$). This score measures how much a layer's resident defenses contribute to mitigating all of the threats in the threat landscape. Let $\mathcal{D}_{\ell} \subset \mathcal{D}$ be the subset of defenses primarily associated with layer $\ell$. The combined coverage ($\kappa_i^{P,\ell}$) against attack $a_i$ using only these defenses is:
\[
\kappa_i^{P,\ell} = 1 - \prod_{j \in \mathcal{D}_{\ell}} \left(1 - \varepsilon_{ij}^P\right)
\]
The resulting LayerScore is then calculated as:
\[
\text{LayerScore}_{\ell}^{P} =
\left(
\frac{
\displaystyle \sum_{i=1}^{39} \tilde{\omega}_i^{P} \kappa_i^{P,\ell}}
{\displaystyle \sum_{i=1}^{39} \tilde{\omega}_i^{P}}
\right) \times 5.
\]
These 7 scores, scaled from 0-5, are presented in a chart (Figure~\ref{fig:safe_r_radar_sub}) to help in the visualization of asymmetric defenses and identification of layers where hardening would be beneficial.

\noindent \textbf{Addressing Methodological Uncertainties with Monte Carlo Simulation:}
 Our assessment inputs—attack likelihoods, impact scores, and defense effectiveness—involve uncertainty arising from historical data, evolving threat intelligence, system-specific details, and expert judgment. We addressed this by running 1,000 Monte Carlo~\cite{monte_calro_application,monte_carlo_usage} iterations of RISK-MAP with input values randomly sampled within ($\pm25\%$)intervals, providing statistically robust measures of platform security despite underlying data uncertainty.

\subsection{Case Studies}
\label{subsec:expt_validation}

\begin{figure*}[h]
    \centering
    \begin{subfigure}[b]{0.42\textwidth}
        \includegraphics[width=\textwidth,trim=0cm 1.5cm 0cm 0.5cm,clip]{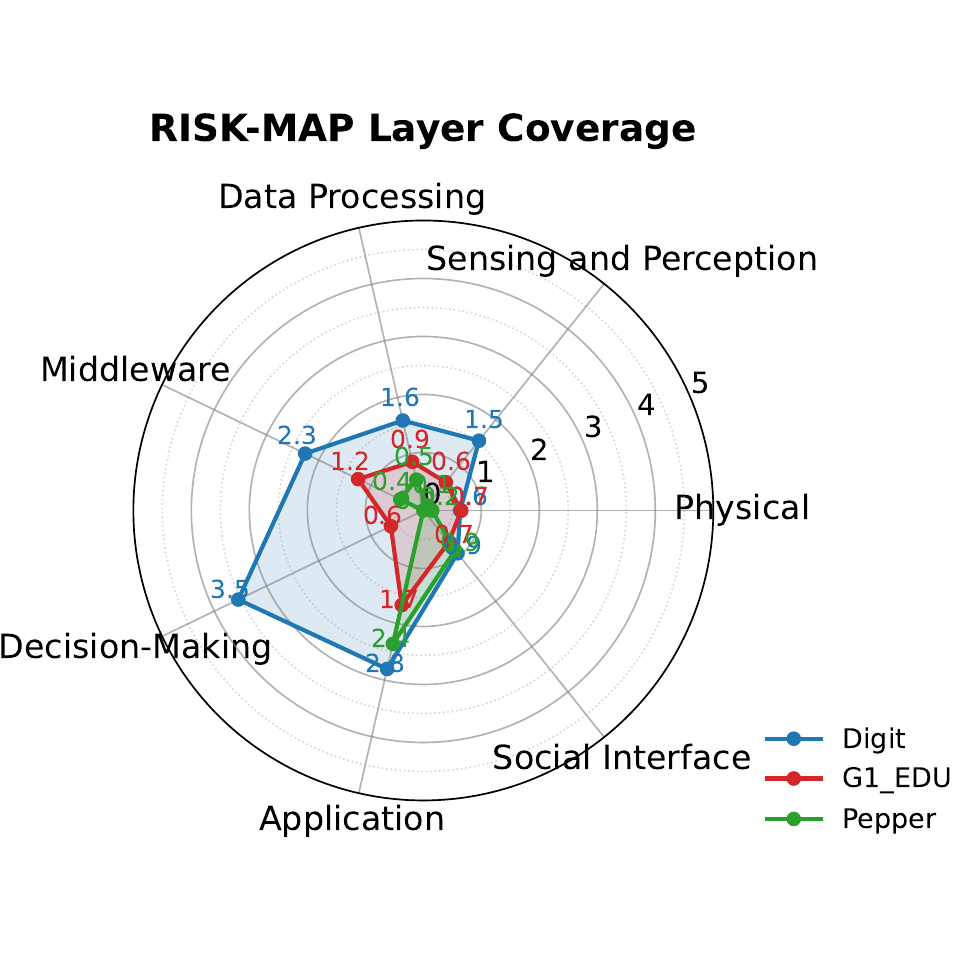}
        \caption{Layer-specific RISK-MAP coverage.}
        \label{fig:safe_r_radar_sub}
    \end{subfigure}
    \hfill
    \begin{subfigure}[b]{0.48\textwidth}
        \includegraphics[width=\textwidth]{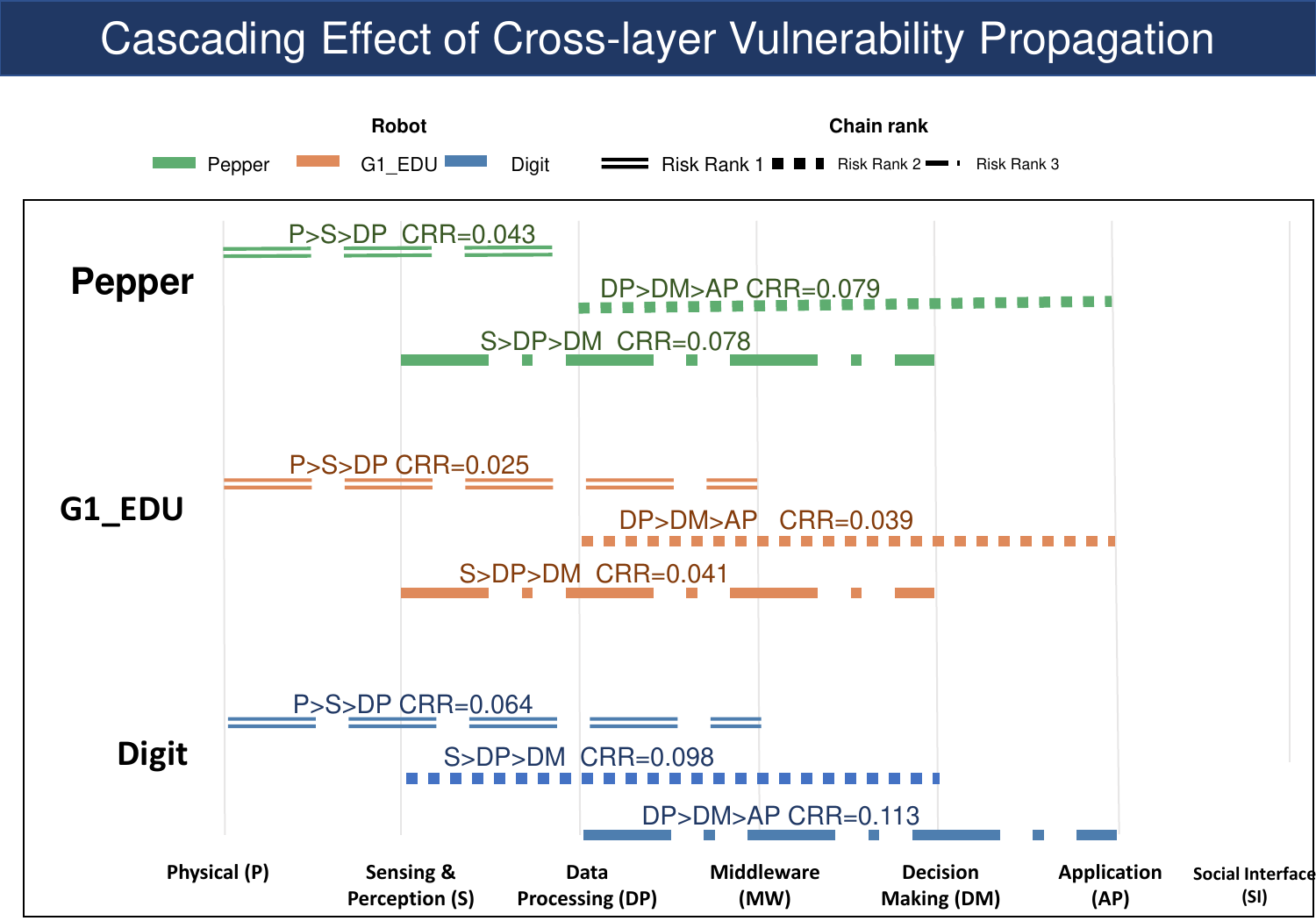}
        \caption{Top-3 cross-layer vulnerability propagation}
        \label{fig:residual_risks_sub}
    \end{subfigure}
    \caption{RISK-MAP results on three humanoids. (a) Layer coverage (seven layers; LayerScore 0–5). (b) Each row shows one robot's highest-risk two-hop attack paths across the autonomy stack (P→SI). Line width encodes Cascade Residual Risk (CRR); line style indicates ranking. Recurring transitions reveal structural attack corridors; thickness variations expose robot-specific coverage gaps.}
    \label{fig:safe_r_combined}
\end{figure*}

To demonstrate RISK-MAP's practical utility, we applied the framework on three representative humanoid platforms: Digit, G1 EDU, and Pepper. 
The assessment was based on publicly available datasheets, official documentation, and a standard deployment scenario. The results of the assessment are presented in Figures~\ref{fig:safe_r_radar_sub},~\ref{fig:residual_risks_sub} and Table~\ref{tab:safer_results_custom}. To assist users in interpretation, we provide multiple complementary views.

\noindent\textbf{Layer diagnostics.} The radar chart (Figure~\ref{fig:safe_r_radar_sub}) provides a layer-specific diagnostic for each platform, highlighting tactical and operational vulnerabilities across each of the three humanoids. This view enables engineering teams to prioritize targeted, subsystem-specific hardening efforts.

    \noindent \textbf{Cross-layer vulnerability propagation. }
    We extend our RISK-MAP assessment to analyze cross-layer vulnerability propagation across our seven-layer model. Using coupling matrices derived from structural feasibility, empirical evidence, and mitigation strength, we enumerate two-hop attack cascades and compute Cascade Residual Risk (CRR) for each robot-path-attack combination. Figure~\ref{fig:residual_risks_sub} reveals systematic attack corridors, with DP→DM and S→DP transitions dominating high-risk paths across all robots, while robot-specific coverage gaps create varying CRR magnitudes for identical attack paths. Detailed methodology and mathematical formulation are provided in Appendix~\ref{appx:cross_layer}.

    \noindent \textbf{Aggregate summary.} Table~\ref{tab:safer_results_custom} summarizes the aggregated scores and key findings for each platform. Digit achieves the highest score (79.5\%), with robust decision-making but remains vulnerable to reward hacking and timing-channel exploitation. G1 EDU (48.9\%) and Pepper (39.9\%) show relative strength in their application layer but are vulnerable to unauthorized API access and network-level replay attacks.

\begin{table}[h]
\caption{RISK-MAP evaluation results.}
\label{tab:safer_results_custom}
\centering
\small
\begin{tabular}{|l|p{5.8cm}|}
\hline
\rowcolor{blue!10} \multicolumn{2}{|c|}{\textbf{Digit (79.5\% ± 3.2\%)}} \\
\hline
Strengths & Decision-making layers \\
\hline
Vulnerabilities & Reward hacking, timing-channel exploitation \\
\hline
\rowcolor{orange!10} \multicolumn{2}{|c|}{\textbf{G1 EDU (48.9\% ± 4.1\%)}} \\
\hline
Strengths & Application layer  \\
\hline
Vulnerabilities & Unauthorized API access, runtime tampering, model-level threats \\
\hline
\rowcolor{green!10} \multicolumn{2}{|c|}{\textbf{Pepper (39.9\% ± 2.8\%)}} \\
\hline
Strengths & Application layer \\
\hline
Vulnerabilities & Replay attacks, denial-of-service, buffer manipulation \\
\hline
\end{tabular}
\end{table}

\noindent\textbf{Methodological scope and results interpretation.}
The aggregated RISK-MAP scores should be interpreted within the defined scope of our assessment method.
Direct comparison of the final RISK-MAP score (\text{RISK-MAP}$_{\%}^{P}$) is most meaningful when the evaluated humanoids possess a similar set of capabilities and therefore, a similar attack surface. The scores are powerful for platform-specific internal assessment and for identifying industry-wide architectural patterns, such as the consistent weakness of the physical layer. However, when comparing humanoids with vastly different features, a less capable platform might achieve a higher score simply by lacking certain vulnerable components. 
An equitable cross-platform benchmark would require normalization against a common capability baseline-a feature not currently implemented in our method. Therefore, the primary value of RISK-MAP lies in its diagnostic capabilities for individual humanoids and its ability to reveal high-level security trends across the humanoid ecosystem, rather than serving as a definitive ranking system for functionally dissimilar platforms.

%% file: conclusion.tex
\section{Summary}
\label{sec:conclusion}

This work addresses critical security challenges as humanoid robots transition to real-world deployment. Our systematic analysis of 39 attacks and 35 defenses across a seven-layer model demonstrates that current humanoid security is fundamentally inadequate for safe human-centered operation.
Our RISK-MAP method reveals systemic vulnerabilities across 3 humanoids. While developers focus on application protections, critical foundational layers remain dangerously exposed. Cross-layer cascade analysis shows attackers can exploit these gaps for privilege escalation and safety bypasses. Monte Carlo validation across 1,000 trials confirms the robustness of our risk measurements.
Humanoid systems exhibit higher vulnerability densities than mature cyber-physical systems, with multiple high-impact attacks cascading across multiple layers. As humanoids deploy in various industries, this security debt threatens public trust and safety. Our coupling matrix analysis reveals that defensive imbalances create systematic bottlenecks where single exploits can cascade into full system compromise. These architectural vulnerabilities require fundamental design changes rather than superficial patches.
The robotics industry must transition from reactive patching to proactive defense architectures. Our methods enables practitioners to identify gaps, prioritize remediation, and validate improvements before deployment. The seven-layer model provides a standardized taxonomy for security assessment across diverse humanoid platforms. Immediate adoption of our methodology by manufacturers could prevent the security debt that has plagued other emerging technologies. As humanoids become ubiquitous in critical infrastructure, the window for establishing robust security foundations is rapidly closing and works such as ours can make the humanoid ecosystems to become safe and secure.

%% file: appendix.tex
\section{Cross-Layer Vulnerability Propagation}
\label{appx:cross_layer}
To understand how security failures spread across humanoid robot systems, we build a $7 \times 7$ matrix $D$ that maps how problems in one layer can affect other layers.

\noindent \textbf{Matrix Construction: }
We model the coupling matrix as:
\begin{equation}
D = (\alpha S + \beta E) \circ (1 - M),
\end{equation}

\begin{itemize}[left=0pt, nosep, leftmargin=*]
  \item $S \in [0,1]^{7 \times 7}$:  How easily problems could spread between layers based on system design,
  \item $E \in [0,1]^{7 \times 7}$: Real-world evidence of problems actually spreading this way,
  \item $M \in [0,1]^{7 \times 7}$: How well current security measures block this spread,
  \item $\alpha, \beta \geq 0$ with $\alpha + \beta = 1$:  How much we trust design analysis ($\alpha$=0.6) vs. real incidents ($\beta$=0.4),
  \item $\circ$: Combines values element by element.
\end{itemize}

We mark each layer as fully connected to itself $D_{ii} = 1$ fbut don't count these self-connections in our analysis.

\noindent \textbf{Avoiding Double Counting}

We separate two types of protection: edge mitigations $M$ that block spread between layers, and layer coverage $C_r(\ell)$ that protects within each layer. This prevents counting the same defense twice.

\noindent \textbf{Finding Vulnerability Paths}

We look for two-step attack paths: $\pi = \ell_i \rightarrow \ell_j \rightarrow \ell_k$ where each hop meets a threshold and the full chain exceeds minimum propagation strength.

\paragraph{Per-hop thresholds:}
\begin{equation}
D_{ij} \geq \varepsilon, \qquad D_{jk} \geq \varepsilon.
\end{equation}

\paragraph{Path strength:}
\begin{equation}
P(\pi) = D_{ij} \cdot D_{jk} \;\;\geq\;\; \text{MIN\_PROP}.
\end{equation}

\noindent \textbf{Parameters}
Unless otherwise stated, we use:
\[
\alpha = 0.6, \quad \beta = 0.4, \quad \varepsilon = 0.3, \quad \text{MIN\_PROP} = 0.1.
\]

We tested longer 3-step paths but found that they didn't change which paths were most dangerous.
\noindent \textbf{Risk Metrics}

\noindent \textbf{Cascade Risk (CRR)}
For robot $r$ and attack $a$, the Cascade Risk along chain $\pi$ indicates residual risk after going through a path and encountering defenses.
\begin{equation}
\CRR(\pi, r, a) = \clip{\, P(\pi) \cdot w_a \cdot U_r(\pi) \, },
\end{equation}
\begin{itemize}[left=0pt, nosep, leftmargin=*]
  \item $w_a$: Attack weight (based on impact, likelihood, detectability),
  \item $U_r(\pi) = \prod_{\ell \in \pi} \bigl(1 - C_r(\ell)\bigr)$: Multiplicative defense gap for robot $r$ along $\pi$,
  \item $C_r(\ell) \in [0,1]$: Per-layer coverage of defenses at layer $\ell$ for robot $r$,
  \item $\clip{x} = \min(1,\max(0,x))$: Clipping function.
\end{itemize}

\noindent \textbf{Cascade Containment Index (CCI)}
The complementary metric indicates how well the system contains the attack.
\begin{equation}
\CCI(\pi, r, a) = 1 - \CRR(\pi, r, a).
\end{equation}

\paragraph{Interpretation.}
\begin{itemize}[left=0pt, nosep, leftmargin=*]
  \item \textbf{High CRR}: Strong propagation paths with high-weight attacks meeting weak defenses.
  \item \textbf{High CCI}: Effective containment through edge mitigations and layer coverage.
\end{itemize}

\noindent \textbf{Uncertainty Analysis}

We assess robustness via Monte Carlo simulation with $1{,}000$ runs, applying i.i.d.\ $\pm 25\%$ uniform noise to:
\begin{itemize}[left=0pt, nosep, leftmargin=*]
  \item $\gamma$ (defense efficacy),
  \item $\mu$ (deployment level),
  \item $w_a$ (attack weights).
\end{itemize}

Results report median values with 5th/95th percentile confidence bands.

\noindent \textbf{Key Findings}

\noindent \textbf{Structural Attack Corridors}
The cascades highlight recurring structural paths, especially $P \!\rightarrow\! S \!\rightarrow\! DP$, $S \!\rightarrow\! DP \!\rightarrow\! DM$ and $DP \!\rightarrow\! DM \!\rightarrow\! AP$ . These corridors consistently rank among the most critical across robots. Their recurrence indicates that vulnerabilities concentrate at these layer junctions. Strengthening these choke-points would yield disproportionate security improvements.

\noindent \textbf{Cross-Robot Risk Heterogeneity}
While the same structural paths appear across \textit{Digit}, \textit{G1\_EDU}, and \textit{Pepper}, their Chain Risk Ratings (CRR) differ noticeably. 
For example, \textit{Digit} shows higher CRRs on \textbf{DP $\rightarrow$ DM $\rightarrow$ AP} (CRR $\approx 0.113$), whereas \textit{Pepper} exhibits comparable risk in \textbf{S $\rightarrow$ DP $\rightarrow$ DM} (CRR $\approx 0.078$). 
This heterogeneity stems from differences in layer coverage and implementation choices, highlighting that robot-specific defenses are necessary even when structural paths overlap.

\noindent \textbf{Attack Path Dominance}
The results confirm that short, high-leverage paths (\textit{h=2}) dominate the Top-3 cascades, often outranking longer chains. 
For instance, \textit{Digit}'s \textbf{DP $\rightarrow$ DM $\rightarrow$ AP} (CRR $=0.113$) is more severe than longer cascades not captured in the Top-3.